\documentclass[12pt]{article}
\usepackage{amsfonts}
\usepackage{booktabs}
\usepackage[T1]{fontenc}
\usepackage[utf8]{inputenc}
\usepackage{bm}
\usepackage{amsmath}
\usepackage{amscd}
\usepackage{amssymb,lscape}
\usepackage[matrix,arrow,curve]{xy}
\usepackage{booktabs}
\allowdisplaybreaks

\numberwithin{equation}{section}       

\newcommand{\bmat}{\left(\begin{array}}
\newcommand{\emat}{\end{array}\right)}

\def\gtrsim{\mathrel{\raise.3ex\hbox{$>$\kern-.75em\lower1ex\hbox{$\sim$}}
}
}

\def\-{\hphantom{-}}

\def\s2{\frac{1}{\sqrt2}}

\def\beq{\begin{equation}}
\def\eeq{\end{equation}}
\def\beqa{\begin{eqnarray}}
\def\eeqa{\end{eqnarray}}

\def\dim{{\rm dim \,}}

\def\eps{\epsilon}

\def\eps{{\epsilon }}

\def\bA{{\bar A}}
\def\bB{{\bar B}}
\def\bC{{\bar C}}
\def\bD{{\bar D}}

\def\mg{m_{3/2}}
\def\mg2{m^2_{3/2}}

\def\Dsl{\,\raise.15ex\hbox{/}\mkern-13.5mu D} 

\def\be{\begin{equation}}
\def\ee{\end{equation}}
\def\bea{\begin{eqnarray}}
\def\eea{\end{eqnarray}}

\DeclareMathOperator{\Es7}{\mathit{E}_{7(7)}}

\newcommand{\nn}{\nonumber}

\begin{document}
\pagestyle{plain}
\begin{titlepage}
\rightline{\small IPhT-T13/055}
\vskip 1.4cm
\begin{center}
\LARGE{Extended geometry and gauged maximal supergravity\\[10mm]}
\large{\bf  G. Aldazabal${}^{a,b}$, M. Gra\~na${}^c$, D.
Marqu\'es${}^d$ , J. A. Rosabal${}^{a,b}$
 \\[4mm]}
\small{
${}^a${\em Centro At\'omico Bariloche,} ${}^b${\em Instituto Balseiro
(CNEA-UNC) and CONICET.} \\[-0.3em]
{\em 8400 S.C. de Bariloche, Argentina.}\\
[0.3cm]
${}^c${\em Institut de Physique Th\'eorique,
CEA/ Saclay \\
91191 Gif-sur-Yvette Cedex, France}  \\
[0.4cm]
${}^d${\em Instituto de Astronom\'ia y F\'isica del Espacio
(CONICET-UBA)} \\
C.C. 67 - Suc. 28, 1428 Buenos Aires, Argentina.  \\
[0.3cm]}
\small{\bf Abstract} \\[0.5cm]

\end{center}

We consider generalized diffeomorphisms on an extended  mega-space
associated to
the U-duality group of gauged maximal supergravity in four dimensions,
$E_{7(7)}$. Through the bein for the extended metric we derive dynamical
(field-dependent) fluxes taking values in the representations allowed by
supersymmetry, and obtain their quadratic constraints from gauge
consistency conditions.
A covariant generalized Ricci tensor is introduced, defined in terms
of a connection for the generalized diffeomorphisms. We show that for
any torsionless and metric-compatible generalized connection, the Ricci
scalar
reproduces   the scalar potential of gauged maximal
supergravity. We comment on how these results extend to other
groups and dimensions.

\vfill

\today

\end{titlepage}

\begin{footnotesize}
\tableofcontents
\end{footnotesize}

\newpage

\section{Introduction}
\label{seci}

The classical symmetry group of the massless fields of string theory is
larger
than that of the metric, namely the group of diffeomorphisms, and includes
purely stringy transformations such as T- or S-duality. The program to
rewrite
the theory in a covariant language under $O(d,d)$, the group that includes
T-duality, goes under the name of Generalized Geometry \cite{Hitchin}, or
Double
Field Theory (DFT) \cite{Hull} (the latter based on previous constructions on Double
Geometry \cite{Hull:2007jy}). In Generalized Geometry, the tangent space,
where the vectors
generating diffeomorphisms live, is enlarged to include the one-forms
corresponding to gauge transformations of the B-field. Instead, in DFT, which aims at providing a field theory approach for strings, the
space itself is doubled, and the
extra half of the coordinates can be thought of as the duals of winding
modes.
Both approaches are related when a section condition is imposed which,
effectively, un-doubles the double space. This condition is sufficient for
consistency of DFT, but only a relaxed version of it is
necessary for gauge consistency at the classical level
\cite{Grana:2012rr}.

By including also S-duality in the game, i.e. by promoting the covariance
to the
full U-duality group, the tangent space or the double space gets enlarged
to an
extended (or exceptional) generalized tangent space  or  a mega-space (a
mega-torus \cite{Dall'Agata:2007sr,Aldazabal:2010ef}, in the case of
toroidal
backgrounds).  The enlarging is such that it accounts for the symmetries
corresponding to RR fields as well as the NSNS, or equivalently combines
the
D-brane charges together with the momentum and winding charges of the
string.
The exceptional tangent space is the starting block of Exceptional 
Generalized Geometry \cite{HullU,Waldram1} (EGG), while we will call the
extension of
DFT to the U-duality groups, first discussed in
\cite{Berman:2011cg}, ``Extended Field Theory'' (EFT). As  DFT,
EFT can be restricted by a section condition
\cite{Coimbra:2011ky}, that also
constraints the fields to depend on a reduced number of physical
coordinates,
but more generally this constraint can also be relaxed.
The U-duality symmetry groups are the exceptional groups $E_{d+1}$ of
toroidal
compactifications, where $d$ is the dimension of the  compactification
space in
string theory (or $d+1$ in M-theory, or rather in 11-dimensional supergravity).

An appealing feature of the extended space is that the stringy
symmetries (diffeomorphisms
plus gauge transformations of all the gauge fields) look just like
diffeomorphisms, and are encoded in a generalized Lie derivative
\cite{Coimbra:2011ky}-\cite{Berman:2012vc}, which gives the differential
structure of
the space. It allows to define a generalized, metric compatible, and
torsion-free connection \cite{Coimbra:2011ky}. Moreover, a
generalized Ricci tensor,  whose flatness condition reproduces the
supergravity
equations of motion, and a generalized Ricci scalar,  encoding the
supergravity
Lagrangian \cite{Coimbra:2011ky}, can be constructed.
For the case of Generalized
Complex Geometry, these tensors were worked out in \cite{Coimbra:2011nw},
inspired from
older double formalism \cite{Siegel:1993th}-\cite{Tseytlin}. More explicit
constructions in DFT are presented in
\cite{Jeon:2011cn}-\cite{hetstring}.

In this paper, we extend the more explicit DFT constructions to the exceptional case,
for  $d=6$, when the symmetry
group is\footnote{All throughout the paper when we refer to $E_7$, we really mean
$E_{7(7)}$.} $E_{7}$, following the definitions introduced
in \cite{Coimbra:2011ky} for EGG. To be more precise, from the generalized Lie
derivative
acting on the bein of the generalized metric we define the ``dynamical
fluxes''. These are required to live in certain representations of the
duality
group.
Jacobi-type constraints on these fluxes are obtained  from requiring
closure of the algebra and gauge invariance. Interestingly enough, for constant fluxes,
which correspond to the embedding tensor of
$N=8$
supergravity \cite{deWit:2002vt},  the  Jacobi-type constraints reduce
to
the usual quadratic constraints of maximal supergravity.
 We show that a generalized Levi-Civita connection, related to the fluxes
by
a torsion free condition, can be constructed allowing to build Riemann and Ricci tensors \cite{Coimbra:2011ky}. When these are defined in the usual way,  they turn out to be non-tensorial, and thus one has to resort to generalized versions of them.
Covariant definitions of these tensors were introduced in \cite{Coimbra:2011ky,Jeon:2011cn,Hohm:2011si}, which interestingly contain undetermined components. Here, extending the definition of
\cite{Jeon:2011cn} to the exceptional case, we find a covariant (though
still
not uniquely defined) version of the generalized Ricci tensor. Taking its
trace,
the undetermined pieces go away, and we show that the generalized Ricci
scalar, which coincides with that of \cite{Coimbra:2011ky} when the section condition is imposed,
can be written purely in terms of the dynamical fluxes. In the case of constant
fluxes, the generalized Ricci scalar is exactly equal to the
potential of
$N=8$ supergravity, provided we identify the generalized metric with the
moduli
space metric for the $N=8$ scalars.
Some definitions and results
for
$d<6$ are given
in \cite{Berman:2012uy}, and some more for $d=4,5$ in
\cite{Musaev:2013rq}. Finally, for completion we provide a list of
complementary results along these lines
\cite{West:2011mm}-\cite{Chatzistavrakidis:2012qj}.

The paper is organized as follows. In  Section \ref{sec: Generalized
diffeomorphisms and fluxes} we set the basic notation and address the
definition
of generalized diffeomorphisms on a $56$ dimensional extended space. The
generalized Lie derivative is introduced and through it the ``dynamical
fluxes'' are defined.
In Section \ref{sec:constraints} we deal with  constraints required from
closure of the algebra of  gauge transformations. We discuss two
different solutions to the Jacobi-type constraints. The so called
\textit{section
condition} (also known as \textit{strong constraint}) and what we call
\textit{twisted
constraints}. We mainly deal with these latter constraints. They reproduce
the quadratic constraints of maximal $N=8$ supergravity and would
naturally appear in Scherk-Schwarz like compactifications.
A geometrical structure for the extended space is addressed  in Section
\ref{sec:A
geometry for the extended space} where a generalized  covariant derivative
and
generalized  torsion are introduced. A generalized Ricci tensor is defined
in
Section \ref{genRiccis}. Interestingly enough it is shown that, under
contractions with the generalized metric, the associated Ricci scalar is
completely determined in terms of generalized fluxes, and corresponds to
the
scalar potential of maximal supergravity.
 Final comments are presented in Section \ref{sec:conlusions}.
Some notation and useful results are summarized in the Appendices.

\section{Generalized diffeomorphisms and fluxes}
\label{sec: Generalized diffeomorphisms and fluxes}

 Our starting point is the ${\bf 56}$-dimensional exceptional generalized
tangent space, or extended space, for an extended version of Double Field
Theory,
where 56 is the dimension of the fundamental representation of
$E_7$.
For toroidal compactifications, such a space was called megatorus
\cite{Dall'Agata:2007sr}\footnote{Here we do not require the extended
space to
be a parallelizable manifold (i.e., a torus or a ``twisted torus"), except for sections \ref{twistedconstraints} and \ref{sec:twistedgenRicci}, and at
the
very end of section \ref{genRiccis}, where we compare the generalized
Ricci
scalar to the potential of $N=8$ supergravity.}. This space encodes all
the
symmetries of string theory compactified on 6-manifolds, or M-theory
compactified on 7-manifolds, namely internal diffeomorphisms and gauge
transformations of the NSNS and RR gauge fields (or the 3-form gauge field
in
M-theory).

The coordinates in the extended space are $Y^M$ with $M = 1,\dots,56$, and the derivatives are
noted by $\partial_M= \partial / \partial Y^M$ and transform in the
fundamental representation of $E_7$. We postpone to Section
\ref{sec:constraints} the discussion about constraints to be satisfied
by them.

We will actually consider an augmented duality group, $\mathbb{R}^+ \times E_7$, which accounts
for a conformal factor \cite{Coimbra:2011ky,Coimbra:2011nw}. This gives an extra degree of
freedom in the 4D supergravity whose string theory origin can be traced to the measure
of the $6$-, $7$-dimensional  manifold,
as well as extra gauge parameters, as we will see.

There is a generalized metric on the extended space $H^{MN}$, which transforms
covariantly under
$\mathbb{R}^+ \times E_7$, and is invariant under  $SU(8)$, the  maximal
compact subgroup of $E_7$. It can be
written in terms of a generalized bein $E_{\bar A}{}^M$
taking values in the quotient $\mathbb{R}^+ \times E_7/SU(8)$
\be
E_{\bar A}{}^M = e^{-\Delta} \tilde E_{\bar A}{}^M\ ,
\ee
where $\tilde E_{\bar A}{}^M$  is an $E_7$ frame, and the conformal factor
$e^{-\Delta}$ corresponds to the component in $\mathbb{R}^+$. In what
follows, the tilde refers to objects that transform under $E_7$ only.
The generalized metric then reads
\beq \label{metricG}
{ H}^{ M  N}=E_{ \bar A}{}^{ M}E_{ \bar B}{}^{ N} H^{ \bar A \bar B}\ ,
\eeq
where $\bar A, \bar B, ... =
1,\dots,56$ are $SU(8)$ planar indices.

The $E_7$ generalized bein $\tilde E_{\bar A}{}^M$ preserves the $Sp(56)$
anti-symmetric matrix $\tilde \omega_{MN}$
\be
\tilde E_{\bar A}{}^M \tilde \omega_{M N} \tilde E_{\bar B}{}^N = \tilde
\omega_{\bar A \bar B} \ , \ \ \ \ \ \tilde \omega_{MN} \tilde \omega^{NP}
= - \delta^P_M\ ,\label{antimetric}
\ee
so it is natural to define a weighted symplectic metric
\be
\omega_{MN}  = e^{2\Delta} \tilde \omega_{MN}\ ,
\ee
which raises and lowers indices according to the following convention
\be
A^M = - \omega^{MN} A_N \ , \ \ \ \ \ \ A_{M} = \omega_{MN} A^N
\label{convention} \ .
\ee
With this convention, the following relations hold
\be
E^{\bar A}{}_M E_{\bar A}{}^N = - \delta^N_M \ , \ \ \ \ \ E^{\bar A}{}_P
E_{\bar B}{}^P = - \delta_{\bar B}^{\bar A} \ , \quad {\rm where}  \
E^{\bar
A}{}_M= -\omega^{\bar A \bar B} \omega_{MN} E_{\bar B}{}^N\ .
\ee
In addition to the symplectic invariant, in $\mathbb{R}^+ \times E_7$ there is also a
quartic invariant
\be
 E^{\bar A}{}_M \ E_{\bar B}{}^N\ E^{\bar C}{}_P \ E_{\bar D}{}^Q\
K^M{}_N{}^P{}_Q =
K^{\bar A}{}_{\bar B}{}^{\bar C}{}_{\bar D} \ .
\label{quarticinv}
\ee

A generalized diffeomorphisms compatible with this symmetry group is
generated by an infinitesimal generalized vector
(or gauge parameter) $\xi$, and is given by the generalized Lie derivative
${\cal L}_\xi$ (or equivalently a generalized gauged transformation
$\delta_\xi$, which coincide when acting on tensors). Acting on a
generalized vector field $V$, we expect it to be a
linear combination of the gauge parameter and its derivatives. A detailed
discussion on how to construct generalized diffeomorphisms can be found in
Appendix \ref{AppC}. Here we simply give its general expression
\be
\delta_\xi V^M ={\cal L}_\xi V^M= \xi^P \partial_P V^M - A^M{}_N{}^P{}_Q
\partial_P \xi^Q V^N + \frac \omega 2 \partial_P \xi^P V^M \
.\label{InternalGenDiffs}
\ee
This was first proposed in \cite{Coimbra:2011ky} and we are using the notation of \cite{Berman:2012vc}
(see  also \cite{Siegel:1993th,Guttenberg:2006zi} in DFT context).
The tensor $A$ is fixed by requiring that the gauge transformations
preserve
the $E_{7}$ structure (see (\ref{antimetric}) and (\ref{quarticinv})) and
is
given by
\be \label{A}
A^M{}_N{}^P{}_Q = 12 \ P_{(adj)}{}^M{}_N{}^P{}_Q = 12 \ (t_\alpha)_N{}^M\
(t^\alpha)_Q{}^P\ ,
\ee
where $t_\alpha$ is a generator of $E_7$, with $\alpha = 1,\dots,133$ an
index
in the adjoint, and $P_{(adj)}$ is a projection to the adjoint $\bf 133$
of
$E_7$. We give its expression in terms of $E_7$ invariants in
(\ref{projector133}). The coefficient $\omega$ corresponds to the
$\mathbb{R}^+$ weight of the object being transformed, which for the bein
$E_{\bar A}{}^M$ is $\omega = 1$, but for $\tilde E_{\bar A}{}^M$ is
$\omega = 0$.\footnote{We use the same symbol $\omega$ for the symplectic
invariant. However, the latter has two indices, while the weight $\omega$
is a
number, so there should be no confusion.}

In the appendices we provide more information about the general structure
of
generalized diffeomorphisms, showing how the $E_7$ case arises as a
particular
example. We also include many useful identities that we will use
repeatedly
along the paper. The relative coefficient between $A$ and the projector
onto the
adjoint depends on the group in question. In Appendix \ref{AppC} we
explain
where this arises from, and why it is $12$ for the case of $\Es7$.

Applying the generalized diffeomorphism (\ref{InternalGenDiffs}) to the
bein,
generated by a bein itself, we get
\be
 \delta_{E_{\bar A}} E_{\bar B} = F_{\bA \bB}{}^{ \bC} E_{\bC}\ ,\label{flux}
\ee
where the ``generalized dynamical fluxes" $F_{\bA \bB}{}^{\bC}$ are
defined as
\be
F_{ \bar A \bar B}{}^{ \bar C} = \Omega_{ \bar A  \bar B}{}^{ \bar C} - 12
P_{(adj)}{}^{\bar C}{}_{\bar B}{}^{\bar D}{}_{\bar E} \Omega_{\bar D \bar
A}{}^{\bar E}+ \frac 1 2 \Omega_{ \bD  \bA}{}^{\bD}
\delta_{\bB}^{\bC}\ ,\label{fluxesplanar}
\ee
where
\be
\Omega_{\bar A \bar B}{}^{\bar C} = E_{\bar A}{}^M \partial_M E_{\bar
B}{}^N
(E^{-1})_N{}^{\bar C} \ . \label{Weizenbockplanar} \
\ee
In the case when there is a global frame (or in other words when the space is parallelizable), this object is called the Weitzenb\"ock connection. Here, with the exception of sections \ref{twistedconstraints}, \ref{sec:twistedgenRicci} and \ref{sec:RicciscalarN-8}, we do not a priori require the existence of such a global frame, and as such our expressions should be understood as local. This means in particular that the generalized dynamical fluxes need not be constant (hence the name dynamical), and furthermore they need not even be globally defined. Nevertheless, by an abuse of notation we will still call these fluxes, and we will call $\Omega$ in (\ref{Weizenbockplanar}) the Weitzenb\"ock connection.

Rotating these expressions with the bein we can define the fluxes with
curved
indices
\be
F_{M  N}{}^{ P} = \Omega_{ M N}{}^{ P} - 12 P_{(adj)}{}^{P}{}_{ N}{}^{
R}{}_{ S}
\Omega_{ R  M}{}^{ S}  + \frac 1 2 \Omega_{RM}{}^R  \delta_{ N}^{P}\ ,
 \label{gaugings}
\ee
and the corresponding Weitzenb\"ock connection in curved indices
\be
\Omega_{M N}{}^{ P} = (E^{-1})_N{}^{\bar B} \partial_M E_{\bar B}{}^P \ .
\label{WeizenbockCurved}
\ee
The Weitzenb\"ock connection (\ref{WeizenbockCurved}) takes values in the
algebra
of $\mathbb{R}^+ \times E_7$  (see Appendix \ref{AppB}), i.e. it can be
written as linear combination
of the generators of $\mathbb{R}^+ \times E_7$
\be
\Omega_{MN}{}^P = - \partial_M\Delta\ \delta_N^P + \tilde \Omega_{MN}{}^P
= \Omega_M{}^0 (t_0)_N{}^P + \tilde \Omega_{M}{}^{\alpha} (t_\alpha)_N{}^P\ ,
\label{WeizAlgebra}
\ee
and is therefore in the $\bf 56 \times \bf (1 + 133)$ product. Here,
$(t_0)_N{}^P = -\delta_N^P$ is the generator of $\mathbb{R}^+$. The $\bf
56 \times 133$ part  \be
\tilde \Omega_{MN}{}^P  = (\tilde E^{-1})_N{}^{\bar B} \partial_M \tilde
E_{\bar B}{}^P \ ,
\ee
contains the irreducible representations ${\bf
56} + {\bf 912} + {\bf 6480}$. The projectors onto the first two
representations
in this product is given by (see Appendix \ref{AppC})
\bea
P_{(56)A}{}^\alpha,^B{}_\beta &=& \frac {56}{133} (t^\alpha t_\beta)_A{}^B
\nn\\
P_{(912)A}{}^\alpha,^B{}_\beta &=& \frac 1 7 \delta^\alpha_\beta
\delta_A^B -
\frac{12}{7} (t_\beta t^\alpha)_A{}^B+ \frac {4}{7} (t^\alpha
t_\beta)_A{}^B\label{projectorsE7} \ .
\eea

Equations (\ref{gaugings}) to (\ref{projectorsE7}) imply that
the fluxes are in the ${\bf 912}$ and ${\bf 56}$ representations only.
More
precisely
\be
F_{AB}{}^C = X_{AB}{}^C + D_{AB}{}^C\ ,\label{embedding tensor}
 \ee
with
\be
X_{AB}{}^C  = \Theta_A{}^\alpha (t_\alpha)_B{}^C \ \ {\rm with}\ \ \ \
\Theta_A{}^\alpha
= 7 P_{(912)A}{}^\alpha,^B{}_\beta\ \tilde \Omega_B{}^\beta\ ,
\label{gaugings912}
 \ee
 and
 \be
 D_{AB}{}^C = - \vartheta_A \delta_B^C + 8 P_{(adj)}{}^C{}_B{}^D{}_A
\vartheta_D  \ , \ \ \ \ \vartheta_A = - \frac 12 (\tilde\Omega_{DA}{}^D -
3 \partial_A \Delta)\ . \label{gaugings 56}
\ee
The fluxes $F$ involve therefore a projection onto the ${\bf 912}$ given
by the gaugings $X$ plus contributions from the gaugings $\vartheta$.  In
the language of gauged supergravity, they correspond to the
gauge group generators, i.e. are  contractions of the embedding tensor
(which dictates how the gauge group is embedded in the global symmetry group) with the
generators of the global symmetry group \cite{deWit:2002vt}. For this reason we
will
sometimes call them ``gaugings''.
The $X$ piece in (\ref{embedding tensor}) corresponds to the $\bf 912$
component
of the fluxes, and in terms of the Weitzenb\"ock connection and the quartic
invariant reads
\be
X_{ABC} = \Omega_{ABC} -  \Omega_{(BC)A}+ 12 K_{BC}{}^{DE} \Omega_{DEA}+
\frac 2
3 \omega_{A(B} \vartheta_{C)} + 8 K_{ABC}{}^D \vartheta_D \ .
\ee
Using the identities (\ref{KK=K}) of the quartic invariant, one can show
that
$X$ enjoys the properties of the $\bf 912$
\be
P_{(adj)}{}^{C}{}_{B}{}^{D}{}_{E}\ X_{A D}{}^{E} = X_{AB}{}^{C}\ , \ \ \
X_{A[BC]}  =
X_{AB}{}^{B} =
X_{(ABC)} =
X_{BA}{}^B  =0\ ,
\ee
which are the well known conditions satisfied by gaugings in $N=8$
maximal supergravity \cite{de Wit:2007mt}.

The $D$ piece (\ref{gaugings 56}) contains two terms, one belonging to the
$\bf 56$ in $\bf 56 \times 1$, and another belonging to the $\bf 56$ in
$\bf 56\times 133$. Notice however, that both terms contain the same
degrees of freedom in terms of $\vartheta$ and are therefore not
independent.

With these results we are able to express the gauge group generators
$(F_A)_B{}^C$ as in \cite{LeDiffon:2008sh}
\be
F_A =  \vartheta_A { t}_0 + (\Theta_A{}^\alpha + 8 \vartheta_B
(t^\alpha)_A{}^B) t_\alpha \ .\label{gaugegenerators}
\ee
The so-called intertwining tensor (i.e., the symmetric components of the
gauge group generators)
takes the form
\be
Z_{AB}{}^C = (F_{(A})_{B)}{}^C =  -
\frac 1 2 (\Theta^{C\alpha} -16 \vartheta_D
(t^{\alpha})^{CD})(t_\alpha)_{AB}\ ,\label{IntertwiningZ}
\ee
and therefore takes values in the $E_7$  algebra, as expected
\cite{LeDiffon:2008sh}.

\section{Consistency constraints}
\label{sec:constraints}

Closure of the algebra  and Leibniz rule of gauge transformations
\be
[{\cal L}_{\xi_1}, {\cal L}_{\xi_2}]\xi_3^M = {\cal L}_{[[\xi_1, \xi_2]]}\xi_3^M={\cal L}_{{\cal L}_{\xi_1}\xi_{2}}\xi_3^M
\ee
impose a set of Jacobi-type constraints on the vector fields, which are quadratic in derivatives.
We first  obtain their general expressions, and then show   two
different set of solutions, commenting on the relevance of each.
Defining
\be
\Delta_{123}{}^M=[{\cal L}_{\xi_1}, {\cal L}_{\xi_2}]\xi_3^M-{\cal L}_{{\cal L}_{\xi_1}\xi_{2}}\xi_3^M=0 \label{deltadef}
\ee
the closure of the algebra  and Leibniz rule can be cast in form
\bea
\Delta_{[12]3}{}^M & = & [{\cal L}_{\xi_1}, {\cal L}_{\xi_2}]\xi_3^M-{\cal L}_{[[\xi_1, \xi_2]]}\xi_3^M\\ \nn
\Delta_{(12)3}{}^M & = &-{\cal L}_{((\xi_1,\xi_2))}\xi_3^M={\cal L}_{[[\xi_1, \xi_2]]}\xi_3^M-{\cal L}_{{\cal L}_{\xi_1}\xi_{2}}\xi_3^M
\eea
where
\be
[[\xi_1, \xi_2]] = \frac{1}{2}({\cal L}_{\xi_1}\xi_2 -{\cal L}_{\xi_2}\xi_1)
\ee
is the Exceptional Courant Bracket
\cite{Waldram1,Coimbra:2011ky},
 while the $((\xi_1,\xi_2))$
is given by
\be
((\xi_1, \xi_2)) = \frac{1}{2}({\cal L}_{\xi_1}\xi_2 +{\cal L}_{\xi_2}\xi_1).
\ee

In the following, it is useful to define the invariant
\cite{Berman:2012vc}
\be \label{Y}
Y^Q{}_M{}^R{}_S =\delta^Q_S \delta^R_M + \frac{1}{2} \delta^Q_M
\delta^R_S - 12 P_{(adj)}{}^Q{}_M{}^R{}_S = \frac{1}{2}  \omega^{QR}
\omega_{MS} - 12 P_{(adj)}{}_{MS}{}^{QR}
\ee
where in the last equality we have used (\ref{idPadj}). The generalized
diffeomorphisms (\ref{InternalGenDiffs}) can be written in terms of this
operator as
\be
{\cal L}_\xi V^M = L_\xi V^M + Y^M{}_N{}^P{}_Q  \partial_P \xi^Q V^N\ee
 where $L_\xi V$ is the ordinary Lie derivative. Therefore, $Y$ can be
understood as an object measuring the departure from usual Riemannian
geometry.

A first constraint arising form closure of the gauge algebra states that
two
successive gauge transformations should effectively correspond to a unique
gauge
transformation. If the first (second) transformation is parameterized by
$\xi_1$ ($\xi_2$), the parameter of the composed transformation is
given by the Exceptional Courant Bracket. Explicitly, this  is
\begin{equation}
[[\xi_1,\xi_2]]^{ M}=\xi_{[1}^{ P}\partial_{ P}\xi_{2]}^M-
A^{ {M}}{}_{ {N}}{}^{ {P}}{}_{ {Q}} \, \partial_{ P} \xi_{[1}^{ {Q}}\,
\xi_{2]}^{ {N}}  + \frac 1 2 \partial_P \xi^P_{[1} \ \xi_{2]}^M\ .
\label{ECB}
 \end{equation}
Using equation (\ref{CyclicPadj}), the closure condition
can be written in the form
\bea
\Delta_{[12]3}{}^M & = &   Y^Q{}_L{}^O{}_I\ \partial_O \xi_{[2}^I\ \xi_{1]}^L\
\partial_Q
\xi_3^M\nn\\
&& +\  A^M{}_N{}^J{}_L Y^Q{}_J{}^O{}_I\ \partial_Q \xi_{[2}^I\ \partial_O
\xi^L_{1]}\ \xi_3^N\nn\\
&& +\  Q^M{}_N{}^{QO}{}_{LI}\ \partial_Q \partial_O\xi_{[2}^I\ \xi_{1]}^L
\
\xi_3^N=0 \label{Closure}
\eea
where we have defined
\be
Q^M{}_N{}^{QO}{}_{LI} = Y^Q{}_J{}^O{}_{(L} A{}^J{}_{I)}{}^M{}_N + \frac 1
2
\omega_{IL} Y^{QMO}{}_N - \frac 1 2 Y^Q{}_L{}^O{}_I \delta^M_N.
\ee
Additionally using
\be
((\xi_1,\xi_2))^M=Y^M{}_N{}^P{}_Q\partial_P\xi_{(1}^Q\xi_{2)}^N
\ee
the Leibniz rule is written as
\bea
-\Delta_{(12)3}{}^M &=& Y^Q{}_L{}^O{}_I \
\partial_Q
\xi_{(1}\  \xi_{2)}^L \ \partial_O \xi_3^M\nn\\
&& -\ Q^M{}_N{}^{QO}{}_{LI} \partial_Q (\xi_{(1}^L \ \partial_O
\xi_{2)}^I)\
\xi_3^N \nn\\
&& -\ \frac 1 4 \omega_{LI}\omega^{QO} \ \partial_Q\xi_1^L \
\partial_O\xi_2^I \
\xi_3^M =0\label{leibnitz1}
\eea

Equations (\ref{Closure}) and (\ref{leibnitz1}) imply that any theory
invariant
under generalized diffeomorphisms will necessarily be a constrained or
restricted
theory, meaning that the fields and gauge parameters will not be generic,
but
must necessarily obey these differential conditions. Therefore, these
constraints select subset of fields and gauge parameters for which the
theory
can be consistently defined.

Notice also that, as shown in \cite{Berman:2012vc}, the Jacobiator can be written as
\be
J(\xi_1, \xi_2,\xi_3) \equiv [[[[\xi_1 , \xi_2]], \xi_3]] + {\rm cyclic} =\frac 1 3 (( [[\xi_1, \xi_2]], \xi_3)) + {\rm
cyclic}
\ee
so even if non-vanishing, it generates trivial gauge transformations by
virtue
of (\ref{leibnitz1}).
We would now like to explore two different set of solutions to the above
constraints, and comment on their relevance for different purposes.

\subsection{The section condition} \label{sec:sec}

In DFT, there exists a so-called {\it section condition}, also known as
the {\it
strong constraint}, consisting of the following differential operator that
must
annihilate all fields and gauge parameters and their products \cite{Hull}
\beq \label{scDFT}
S=\eta^{mn}\partial_m \partial_n \ ,
\eeq
where $m, n=1,...,2d$ span the fundamental of $O(d,d)$. In the strong
version of
the constraint, each partial derivative acts on a given field. In its weak
version, what should vanish is the second order operator acting on a
single
field.
The result of the strong constraint is that the fields and gauge
parameters no
longer depend of the full set of $2d$ coordinates, but rather on a
$d$-dimensional section of the space.

The $E_7$ version of the constraint is given by the following operator \cite{Coimbra:2011ky}
\beq \label{scE7}
S_{MN}=P_{(adj)}{}_{MN}{}^{QR} \partial_Q \partial_R \ ,
\eeq
and again in its weak version this whole operator should vanish when
acting on a
single field, while in its strong version each derivative hits a different
field.

In $SL(8)$ indices\footnote{The decomposition of $E_7$ representations
into
$SL(8)$ ones is the same as for $SU(8)$ representations, and these are
given in
Appendix \ref{AppA}} the ${\bf 63}$ and ${\bf 70}$ pieces of the strong
version
of the constraint $S_M{}^N=0$ read (see equation (\ref{56x56=133SLE}))
\bea
0&=&\partial^{cb} A \, \partial_{ca} B -\frac{1}{8} \delta_a^b \,
\partial^{cd}
A \, \partial_{cd} B+ \partial^{cb} B \, \partial_{ca} A -\frac{1}{8}
\delta_a^b
\, \partial^{cd} B \, \partial_{cd} A \nn \\
0&=& \partial_{[ab} A \, \partial_{cd]} B + \frac{1}{4!}
\epsilon_{abcdefgh} \,
\partial^{ef} A \, \partial^{gh} B\ ,
\eea
for any pair of fields $A, B$, where $a,b,...=1,..,8$ are fundamental
$SL(8)$
indices. It is not hard to see that implies that the fields can only
depend on
$7$ out of the $56$ coordinates of the extended space
\cite{Berman:2012vc}.
Indeed, calling these directions $\hat a=1,...,7$, we get that, up to
$E_7$
rotations,  $\partial_{\hat a 8}\equiv\partial_{\hat a}, \partial_{\hat a
\hat
b}=0, \partial^{ab}=0$ is the only solution to the constraint. This is
precisely
how the derivative, which spans a 7-dimensional space for
compactifications of
M-theory on 7-dimensional manifolds, is embedded in the fundamental
representation of $E_7$ in Exceptional Generalized Geometry
\cite{Waldram1}.

To make contact with DFT, or equivalently with
compactifications
of type II theories, it is useful to use the $SL(2) \times O(6,6)$
subgroup of
$E_7$, under which the fundamental representation breaks according to
$M=(\hat
\imath m, \alpha)$ where $\hat \imath=+,-$ ($m=1,...,12$) is a fundamental
of
$SL(2)$ ($O(6,6)$), and $\alpha$ is a spinorial index in the ${\bf 32}$ of
$O(6,6)$.
The $56$ coordinates of the extended space contain therefore two copies of
the
double torus in DFT $\partial_{\pm m}$ (which are related by an $SL(2)$
S-duality) plus 32 spinorial directions $\partial_{\alpha}$.

The different components of the projector onto the adjoint representation
are
given in (\ref{adjprojTsub}). In particular, the first one of them gives
the
following constraint
\beq \label{Scomp}
S_{\hat \imath m \hat \jmath n}=- \tfrac{1}{2} \, \epsilon_{\hat \imath
\hat
\jmath} \, \partial_{[+|m} A \, \partial_{|-]n} B + \tfrac{1}{12} \eta_{mn}
\partial_{(\hat \imath|p} A\,  \partial_{|\hat \jmath) q} B \, \eta^{pq}  -
\tfrac{1}{8} \,
\eps_{\hat \imath \hat \jmath} \, \partial_{\alpha} A\left[ \gamma_{mn}
\right]^{\alpha \beta} \partial_{\beta} B \ .
\eeq
The requirements $S_{+m+n}=0$, $S_{-m-n}=0$ give on one hand that the
 derivative can only span one of the two
copies of the DFT coordinates (call it $+$, or {\it electric}), and on the other hand
we get exactly the DFT strong constraint
(\ref{scDFT}), which implies that the dependence is only on half of the
double coordinates, that we call $\overline{m} = 1, \dots,6$.
Using this in the $S_{+m-n}-S_{-m+n}$ constraint, only the last term in (\ref{Scomp}) survives, and one can see that the
spinorial derivative can only span a two-dimensional (12-dimensional)
subspace
in the case of positive (negative) chirality. The former is relevant for compactifications of
type IIA, where D-branes
have
even dimensionality, while  the latter applies to type
IIB. The allowed components of the spinorial
derivatives are further reduced by the condition $S_{+m\alpha}=0$
(where
the relevant projector is given in the last line of (\ref{adjprojTsub}))
to a
one-dimensional (zero-dimensional) space in the case of positive
(negative)
chirality, defined by the constraint
 $[\gamma^{\overline{m}}]_\alpha{}^\beta \partial_{\beta}=0$\footnote{In
the
one-to-one correspondence between O(6,6) spinors and forms on the
cotangent
space $T^*_6 M$, this corresponds to the six-forms.}. In summary, the
strong
constraint implies that the dependence is on 6 coordinates
only\footnote{This is
precisely how the derivative is embedded into an $E_7$ representation for
compactifications of type II theories in EGG \cite{Grana:2009im}.}
$\partial_{+\overline{m}}\neq 0 $ , plus an extra spinorial coordinate for
the
case of positive chirality, which arises in compactifications of type IIA.
This
extra coordinate is nothing but the M-theory circle. If one wants to avoid
this
dependence, an extra constraint should be supplemented to (\ref{scE7}).

With this information, we can now show that a possibility to solve the
constraints (\ref{Closure}) and (\ref{leibnitz1}) is to restrict the
vector
fields to depend only on a 7-dimensional slice of the full mega-space,
i.e.
constrain them to satisfy the section condition
\be
P_{(adj)MN}{}^{PQ} \ \partial_P V^R\ \partial_Q U^S = 0 \ \ \ \ \ \ \
P_{(adj)MN}{}^{PQ} \ \partial_P \partial_Q U^S = 0\ .
\label{sectioncondition}
\ee
Notice that in (\ref{Closure}) and (\ref{leibnitz1}) all the derivatives
are
contracted with the $Y$ tensor defined in (\ref{Y}). When acting on two
partial
derivatives $\partial_Q \partial_R$, the last term on the last equality in
(\ref{Y}) vanishes due to the section condition. The first term vanishes
because
the section condition restricts the derivatives to lie in an isotropic
bundle,
i.e. to have zero inner product with respect to the symplectic form.
Therefore  the section
condition implies
\beq \label{sectionY}
Y^Q{}_M{}^R{}_S \, \partial_Q \partial_R =0 \ ,
\eeq
acting on any two fields, and this in turn guarantees that (\ref{Closure}) and (\ref{leibnitz1})
are satisfied.

\subsection{Twisted constraints}
\label{twistedconstraints}

In \cite{Grana:2012rr}
it was
shown that the section  condition is sufficient for the gauge consistency
of the
theory, but not necessary. Rather, in particular setups such as the ones discussed below, it is
not hard to see that the constraints (\ref{Closure}) and (\ref{leibnitz1}) are weaker than the
strong constraint, and this allows for extended configurations not
solving the section condition.

These weaker conditions apply when the vector fields are required to take the
form
\be
V^M  = E_{\bar A}{}^M (Y)\  v^{\bar A} \label{vectors}
\ee
with $v$ constant. Note that this is only possible in parallelizable manifolds, so in this section (as well as in section \ref{sec:twistedgenRicci}), where we use these weaker constraints, we restrict to this case (i.e. we demand that the extended space is parallelizable). This definition can be trivially extended to tensors
with
more indices. This kind of behavior arises naturally in the context of
Scherk-Schwarz compactifications, where the bein plays the role of a twist, and the $v^{\bar A}$ vectors
correspond to
fields or gauge parameters in the effective action, and therefore only
depend on
space-time coordinates (which we are ignoring in this paper, thus $v$ is
constant here). We therefore call the constraints obtained in thee setups ``twisted constraints''.
We will show that requiring the vector fields in planar indices to be constant,
equations (\ref{Closure}) and (\ref{leibnitz1}) admit solutions that do not necessarily satisfy
the section condition.

Let us now move to closure (\ref{Closure}) and the Leibniz rule
(\ref{leibnitz1}).
Notice that in the particular case when (\ref{Closure}) is evaluated on frames,
 i.e. $\xi_1\rightarrow E_{\bar A}$,
$\xi_2\rightarrow E_{\bar B}$ and $\xi_3\rightarrow E_{\bar C}$
\bea\nn
\Delta_{[\bar A \bar B] \bar C}{}^{M} & = & Y^Q{}_L{}^O{}_I\ \partial_O E_{[\bar B}{}^I\ E_{\bar A]}{}^L
\partial_Q E_{\bar C}{}^M\\\nn
{} & {} & \ + A^M{}_N{}^J{}_L Y^Q{}_J{}^O{}_I\ \partial_Q E_{[\bar B}{}^I\ \partial_O
E_{\bar A]}{}^L\ E_{\bar C}{}^N\nn\\
{} & {} & +Q^M{}_N{}^{QO}{}_{LI}\ \partial_Q \partial_OE_{[\bar B}{}^I\ E_{\bar A]}{}^L E_{\bar C}{}^N\label{ClosureE}
\eea
and the same for (\ref{leibnitz1})
\be
{\cal L}_{((E_{\bar A}, E_{\bar B}))}E_{\bar C}{}^M = -\Delta_{(\bar A\bar B)\bar C}{}^{M}.
\label{Leibnitz11}
\ee

Alternatively one can compute $\Delta_{\bar A\bar B\bar C}{}^{M}$ using (\ref{flux})
directly in  terms of $F_{\bar A\bar B}{}^{\bar C}$
\bea \label{Delta4}
\Delta_{\bar  A\bar B\bar C}{}^{\bar D} &=&  \left([F_{\bar A},F_{\bar B}]
+ F_{\bar A\bar B}{}^{\bar E}F_{\bar E}\right){}_{\bar C}{}^{\bar D}\\
&& + 2 {\partial}_{[\bar A} F_{\bar B]}{}_{\bar C}{}^{\bar D} + A^{\bar
D}{}_{\bar C}{}^{\bar E}{}_{\bar F} {\partial}_{\bar E} F_{\bar
AB}{}^{\bar F} - \frac
12 {\partial}_{\bar E} F_{\bar A\bar B}{}^{\bar E} \delta_{\bar C}^{\bar
D}. \nn
\eea
Then the closure and the Leibniz rule are recovered in the form
\be
\Delta_{\bar  A\bar B\bar C}{}^{M}=0. \label{Delta5}
\ee
Notice that if the section condition is imposed the closure and the Leibniz rule are guaranteed
as noted above. However if we restrict our vectors to take the form (\ref{vectors}) a weaker
version of the constraints can be considered provided the quadratic constraint  (\ref{Delta4})
is imposed to ensure the consistency of the theory.
In the particular case in which the fluxes are constant, condition
(\ref{Delta4})  becomes
\be
[F_{A}, F_{B}] = - F_{AB}{}^{C}F_{C}
\ee
which are precisely the quadratic constraints of maximal supergravity,
even  with local
scaling symmetry \cite{LeDiffon:2008sh}(notice that the trace of this
equation implies in turn $F_{AB}{}^{C}\vartheta_{C} = 0$).

Let us mention that when the
fluxes
are expressed in planar indices, they should transform as scalars. In
fact, we
find
\be \label{Deltadefinition}
\delta_\xi F_{\bar N\bar P}{}^{\bar M} = \xi^{\bar Q}{\partial}_{\bar Q}
F_{
\bar N\bar P}{}^{\bar M} - \xi^{\bar Q} \Delta_{\bar Q \bar N\bar P
}{}^{\bar M}
\ee
This actually guarantees  that $X$ and $\vartheta$ transform as scalars
independently. For example, in the case of $\vartheta$ we get
\be
\delta_\xi \vartheta_{\bar M} = \xi^{\bar N} {\partial}_{\bar N}
\vartheta_{\bar
M}  - \frac 1 {56}\xi^{\bar N} \Delta_{\bar N \bar M \bar P}{}^{\bar P}.
\ee

In summary, the constraints (\ref{Delta5}) are the only
necessary and sufficient conditions for consistency of the theory at the
classical level when the vectors are restricted as in (\ref{vectors}).
So, while the relaxed constraints are necessary and sufficient for gauge
consistency in this case, the section condition is only sufficient.
In DFT, explicit examples of truly extended configurations were
found
in \cite{Dibitetto:2012rk}, and it would be interesting to find some here
as
well.

A few words are in order. The section condition is crucial to make contact
with
10 or 11-dimensional supergravity, and therefore puts the extended theory
in a
safe and controlled place. When relaxed, the connection between this
construction and higher dimensional supergravity is less clear and should
be
understood better. Configurations that satisfy the relaxed constraints but
not
the section condition lie beyond supergravity compactifications, and are
therefore strictly non-geometric (we refer to \cite{Dibitetto:2012rk} for
a
discussion on these issues). Whether they correspond to allowed
configurations
in the full string or M-theory is a question that remains partially unanswered and
worth
exploring.

\section{A geometry for the extended space}
\label{sec:A geometry for the extended space}

In this section we discuss the covariant derivative on the extended
space, seek
for a covariant definition of torsion, and propose a set of conditions
that the
connection must satisfy. Later in the following section we will define a generalized
Ricci tensor, and show
that under contractions with the generalized metric, the associated Ricci
scalar
is completely determined in terms of generalized fluxes, and corresponds
to the
scalar potential of maximal supergravity.

Having defined the generalized notion of Lie derivative in
(\ref{InternalGenDiffs}), it is natural to look for derivatives that
behave
covariantly under such  transformations. We begin by defining the
covariant
derivative of a bein $E$ as
\be
\nabla_M E_{\bar A}{}^N = W_{M\bar A}{}^{\bar B} E_{\bar B}{}^N=
\partial_M
E_{\bar A}{}^N  + \Gamma_{MP}{}^N E_{\bar A}{}^P \ ,
\ee
in terms of a Christoffel connection $\Gamma$, or alternatively a spin
connection
$W$. They are related to the Weitzenb\"ock connection $\Omega_{\bar A}$
(\ref{Weizenbockplanar}) taking values in the algebra of $\mathbb{R}^+
\times E_7$ (see Appendix
\ref{AppB}). The three connections are related through
\be
W_{ C A}{}^{  B} = \Omega_{  C   A}{}^{  B} + \Gamma_{  C
 A}{}^{ B}\ .\label{relatedconnections}
\ee
In addition, one can relate the gaugings to the Weitzenb\"ock
connection
through projections, as in equation (\ref{fluxesplanar}).

These connections must also transform properly so as to compensate the
failure
of the usual derivative to transform as a tensor. Given that the covariant
derivative is requested to transform covariantly, so must the spin
connection.
Hence, taking into account (\ref{relatedconnections}), we see that the
Christoffel connection must fail to transform as minus the failure of the
Weitzenb\"ock connection
\be
\Delta_\xi \Omega_{ B   C}{}^{  D} = - \Delta_\xi \Gamma_{  B
C}{}^{  D}\ ,
\ee
where $\Delta_\xi = \delta_\xi - {\cal L}_\xi$ is defined as  in an
analogous way as in (\ref{Deltadefinition}), and
represents the failure of an object to transform as a tensor.

\subsection{Generalized connections and torsion}

We can define the generalized torsion through  \cite{Coimbra:2011nw}
\be
{\cal T}_{\bar A \bar B}{}^{\bar C} \equiv (E^{-1})_M{}^{\bar C} ({\cal
L}^{\nabla}_{E_{\bar A}}-{\cal L}_{E_{\bar
A}})  E_{\bar B}{}^M\ ,
\label{torsion}
\ee
 where ${\cal L}^{\nabla}$ is defined as in (\ref{InternalGenDiffs}), but
with a
partial replaced by a covariant derivative. Using
(\ref{relatedconnections}) we arrive at
\be \label{torsionGamma}
{\cal T}_{  A   B}{}^{  C} = \Gamma_{AB}{}^C - 12
P_{(adj)}{}^C{}_B{}^P{}_Q \Gamma_{PA}{}^Q+\frac 1 2
\Gamma_{D A}{}^D \delta^C_B \ .
\ee
A
torsionless connection requires this to vanish, which amounts to
\be
\Gamma_{  A   B}{}^{  C} = 12 P_{(adj)}{}^{  C}{}_{  B}{}^{  D}{}_{  E}
\Gamma_{
 D   A}{}^{  E} - \frac 1 2 \delta^{  C}_{  B} \Gamma_{  D   A}{}^{
D}\ .\label{torsionless}
\ee
Now, acting on this condition with the projector to the adjoint in the
last two
indices, we find
\be
P_{(adj)}{}^{ G}{}_{  H}{}^{  B}{}_{  C} \Gamma_{  A   B}{}^{  C} = 12
P_{(adj)}{}^{  G}{}_{  H}{}^{  B}{}_{  C}\ \Gamma_{  B   A}{}^{  C}\ ,
\ee
and now plugging this result in (\ref{torsionless}) we find
\be
\Gamma_{  A   B}{}^{  C} = P_{(adj)}{}^{  C}{}_{  B}{}^{  D}{}_{  E}
\Gamma_{  A
  E}{}^{  D} - \frac 1 2 \Gamma_{  D   A}{}^{  D} \delta_{  B}^{  C}\ ,
\label{GammaProj}
\ee
so the trace of the connection  measures its failure  to take values in
the $\bf 56 \times 133$ of $E_7$. Notice the two traces of the connection are
related
\be
\Gamma_{AB}{}^{B}=-28\Gamma_{BA}{}^{B}.\label{twotraces}
\ee

Let us now dedicate a few lines to comment on the relation between the
trace of the Christoffel
connection and the $\mathbb{R}^+ \times E_7$ measure. Notice that $\sqrt{
H }$ does not transform
as a density under the generalized diffeomorphisms
(\ref{InternalGenDiffs}), instead the proper
measure is given by $(\sqrt{H})^{-1/28} = e^{-2\Delta}$ since
\be
\delta_\xi e^{-2\Delta} = \partial_P (e^{-2\Delta} \xi^P)\ .\label{measure}
\ee
Partial integration of the covariant derivative in the
presence of the $\mathbb{R}^+ \times E_7$ density   $e^{-2\Delta}$
\be
\int e^{-2\Delta} U \nabla_M V^M = - \int e^{-2\Delta} V^M \nabla_M U \  \
\ \ee
is ensured if
\be
\Gamma_{MN}{}^M = - 2 \partial_N \Delta \ .\label{trace}
\ee
which, as we show below, is a consequence of metric compatibility.
The torsionless (\ref{torsionless}) and the trace equations (\ref{trace})
imply that the connection takes the form
\be
\Gamma_{MN}{}^P = \tilde \Gamma_{MN}{}^P - \frac {24}{19}
P_{(adj)}{}^P{}_N{}^K{}_M \partial_K \Delta + \partial_M \Delta \delta_N^P\ ,
\label{Gamma}
\ee
where $\tilde \Gamma$ belongs to the $\bf 6480$ of $\bf 56 \times 133$
(which is traceless).
This piece remains undetermined here, but a part of it will be fixed by imposing metric
compatibility.

We now turn to the analysis of the spin connection $W$. Using
(\ref{relatedconnections}),
(\ref{WeizAlgebra}) and (\ref{Gamma}) one can obtain
\be
W_{AB}{}^C = \tilde \Omega_{AB}{}^C + \tilde \Gamma_{AB}{}^C - \frac
{24}{19} P_{(adj)}{}^C{}_B{}^P{}_A \partial_P \Delta = \left(\tilde
\Omega_A{}^\alpha + \tilde \Gamma_{A}{}^\alpha - \frac{24}{19}
(t^\alpha)_A{}^P \partial_P \Delta\right) (t_{\alpha})_B{}^C\ .
\ee
From this expression we learn on the one hand that the spin connection
takes values in $\bf 56 \times 133$, and on the other that its trace is
proportional to the $\vartheta$ gaugings  (\ref{gaugings})
\be
W_{AB}{}^A = - 2 \vartheta_B\ .\label{spintrace}
\ee
Knowing that the spin connection belongs to the $\bf 56 \times 133$, we
can now act with the projectors to its irreducible representations, and
find
\bea
P_{(912)AB}{}^C,^{MN}{}_P W_{MN}{}^P &=& \frac 1 7 X_{AB}{}^C\nn\\
P_{(56)AB}{}^C,^{MN}{}_P W_{MN}{}^P &=& - \frac {16}{19}
P_{(adj)}{}^C{}_B{}^D{}_A \vartheta_D \ , \label{projsSpin}
\eea
while the projector to the $\bf 6480$ just relates it to $\tilde \Gamma$
which is undetermined by the torsionless condition.
Finally let us write explicitly the expression for the torsionless
condition in terms of $W$, because it will be useful in the following
\be
W_{AB}{}^C - 12 P_{(adj)}{}^C{}_B{}^D{}_E \ W_{DA}{}^E = F_{AB}{}^C +
\vartheta_A \delta_B^C
\label{spincon912} \ .
\ee

\subsection{Generalized metric compatibility}
Another condition we can impose to the connections is  compatibility with
the
generalized metric $H$, equation (\ref{metricG}). The constraint reads
\be
0 = \nabla_A {H}^{BC} \ \ \ \ \Rightarrow \ \ \ \   W_{A}{}^\alpha
(t_\alpha)_{\bar D}{}^{(\bar B} {H}^{\bar C) \bar D} = 0\ .
\label{metriccomp}
\ee

It is instructive to turn to the $SU(8)$ language. When $E_7$ is broken to
$SU(8)$, the fundamental $\bf 56$ and adjoint $\bf 133$ representations
break
according to
\be
M = {\bf 56} = {\bf 28} + \overline{\bf 28} = (M_{[ij]},M^{[ij]})\ , \ \ \ \
\ \
\alpha = {\bf 133} = {\bf 63} + {\bf 70} = (\alpha_i{}^j,\alpha_{[ijkl]})\ ,
\ee
where $i,j,\dots = 1,\dots,8$ and the $\bf 63$ is traceless $\alpha_i{}^i = 0$.
One can then see that metric compatibility equations (\ref{metriccomp})
reduce
to
 \be
 W_A{}^{[ijkl]} = 0\ .\label{con70}
 \ee
 Hence the projection of the spin connection to the $\bf 70$ of $\bf 133$
must
vanish. On the other hand, metric compatibility leaves the $\bf 63$ piece
of the
connection undetermined.

Another consequence of the metric compatibility
is Eq. (\ref{trace}). This can be proven taking into account
(\ref{twotraces}), the fact that the metric is covanriantly constant (Eq. (\ref{metriccomp}))
and the useful formula for the derivative of the determinat
\be
\partial_AH=-HH_{BC}\partial_AH^{BC} \ , \qquad H\equiv{\rm det} H_{AB}=e^{2\cdotp56\Delta}
\ee

Let us also comment on compatibility with $E_7$ invariants. The
compatibility
with the symplectic metric is not an additional constraint on the
connection,
but is automatically satisfied, in fact
\be \label{compsympl}
\nabla_M \omega^{PQ} = - 2 \partial_M \Delta\ \omega^{PQ} - 2
\Gamma_M{}^{[PQ]} = 0\ ,
\ee
where we have used (\ref{Gamma}). The same holds for the compatibility
with the quartic invariant, since
\be
\nabla_M K^{PQRS} = - 4 \partial_M \Delta\ K^{PQRS} +  4
\Gamma_{ML}{}^{(P}K^{QRS)L}  = 0\ ,\ee
is automatically satisfied using (\ref{propsE7}).

We summarize the properties of the different connections introduced, as
well as
that of the fluxes, in Table 1.

\begin{table}
\begin{center}
\scriptsize {\begin{tabular}{| l | c | c | c| c|}\hline
& $\Omega$ & $F$ & $W$ & $\Gamma$ \\
\hline
Name & Weitzenb\"ock  & Fluxes & Spin  & Levi-Civita    \\
\hline
Definition & $(E^{-1})_N{}^{\bar A} \partial_M E_{\bar A}{}^P$ & $\Omega_{
M N}{}^{ P}
- A^{P}{}_{ N}{}^{ R}{}_{ S} \Omega_{ R  M}{}^{ S} + \frac 1 2
\Omega_{QM}{}^Q \delta_N^P$
& $\Omega + \Gamma $ & $\nabla-\partial$
 \\ \hline
$\mathbb{R}^+ \times E_7$ Rep & ${\bf 56} + {\bf 912}+{\bf 6480}$ & ${\bf
56} (D) + {\bf
912} (X) $ & $({\bf 56}+ {\bf 912})(F)$ & ${\bf 56} + {\bf 6480} (\tilde
\Gamma)$
\\
& & & + ${\bf 6480}$ & \\
\hline
 Undetermined & $----$ & $----$  & \multicolumn{1}{c}{\,\,\,\,\,\,\,\,\,\,\,\
${\bf 6480}$} &
\\
 (torsionless) & & & \multicolumn{1}{c}{}&  \\
 \hline
 Undetermined& $----$ & $----$
&\multicolumn{1}{c}{\,\,\,\,\,\,\,\,\,\,\,\,\,\,\,\,  ${\bf 56\times 63}$}
&
 \\
 (Metric comp.) & & & \multicolumn{1}{c}{}&  \\ \hline
\end{tabular}}
\caption{\small
Definitions and properties of the different connections introduced. In
parenthesis we have indicated the name given to the particular
representations. ``Undetermined (torsionless/Meric comp.)'' means that the given component is not fixed by the torsionless/metric compatibility condition.}
\end{center}
\label{ta:connections}
\end{table}

\section{Generalized Ricci tensor and the scalar potential of gauged maximal
supergravity} \label{genRiccis}

The Riemann  and torsion tensors are usually defined through the relation
\be
[\nabla_M , \nabla_N] V_P = - R_{MNP}{}^L V_L - T_{MN}{}^L \nabla_L V_P\ ,
\ee
with
\be
R_{MNP}{}^R = \partial_M \Gamma_{NP}{}^R - \partial_N \Gamma_{MP}{}^R +
\Gamma_{ML}{}^R \Gamma_{NP}{}^L - \Gamma_{NL}{}^R \Gamma_{MP}{}^L\ ,
\label{Riemann}
\ee
and
\be
T_{MN}{}^P = \Gamma_{MN}{}^P - \Gamma_{NM}{}^P\ .
\ee

We already discussed a generalized version of torsion, arguing that the usual definition is
non-covariant under generalized diffeomorphisms (\ref{InternalGenDiffs}). The same happens to
the Riemann tensor and its trace (the Ricci tensor), and then one has to resort to generalized
versions of them. We will now split the discussion in two parts. We will begin
with
the definition of a generalized Ricci tensor, that is covariant under
generalized diffeomorphisms that close under the section condition. Then, we will
extend the
definition of this tensor so that it is also covariant under generalized diffeomorphisms that
close under the twisted constraints.

\subsection{Generalized Ricci tensor and the section condition}
In this section we will restrict to diffeomorphisms parameterized by
vectors
obeying the section condition, defined in Section \ref{sec:sec}.

As we mentioned, under generalized diffeomorphisms
(\ref{InternalGenDiffs}), the
generalized Riemann tensor is not covariant. Using the section condition
\eqref{scE7} (or its variant \eqref{sectionY}), we find
\be
\Delta_\xi R_{MNK}{}^L = 2 \Delta_\xi \Gamma_{[MN]}{}^Q \Gamma_{QK}{}^L \
, \ \
\ \ \ \Delta_\xi \Gamma_{MP}{}^N
 = 12 P_{(adj)}{}^N{}_P{}^R{}_S \partial_M \partial_R\xi^S - \frac 1 2
\partial_M \partial_Q \xi^Q  \delta_N^P\ ,\ee
which in the usual case (i.e. in ordinary general relativity where $Y = 0$) vanishes due
to
vanishing torsion. Notice that for everything to be consistent here, we must have  $\Delta_\xi
\Gamma_{NM}{}^N = -2 \Delta_\xi (\partial_M \Delta)$, and this holds up to terms that
vanish under the section condition.

The usual Ricci tensor is defined as
\be \label{Ricci}
R_{MN} = R_{MPN}{}^P\ ,
\ee
and in this case is not symmetric, and fails to transform covariantly as
\be
\Delta_\xi R_{MN} = 2 \Delta_\xi \Gamma_{[MQ]}{}^P \Gamma_{PN}{}^Q\ .
\ee
However, note that the vanishing (generalized) torsion condition
(\ref{torsionless}) imposes
\be
2 \Gamma_{[MN]}{}^Q = - Y^Q{}_N{}^R{}_P \Gamma_{RM}{}^P\ ,
\ee
and this allows to rewrite
\be
2 \Delta_\xi \Gamma_{[MP]}{}^Q \Gamma_{QN}{}^P = 2\Delta_\xi
\Gamma_{PM}{}^Q
\Gamma_{[NQ]}{}^P\ .
\ee
Using this, is it easy to see that the following symmetric object
\be
{\cal R}_{MN} \equiv \frac 1 2 \left( R_{MN} + R_{NM} + \Gamma_{R M}{}^P\
Y^R{}_P{}^S{}_Q \ \Gamma_{SN}{}^Q\right) = {\cal R}_{NM}\ ,
\label{generalizedRicciTensor}
\ee
is a  covariant extension of the Ricci tensor
\be
\Delta_\xi {\cal R}_{MP} = 0 \ .
\ee
This is the natural
extension of the DFT definition of Ricci tensor introduced in
\cite{Jeon:2011cn,Hohm:2011si}.

Let us conclude this section by noticing that a
definition of the generalized Ricci
tensor can be given in terms of covariant derivatives. In fact,
after some algebra we find that \cite{Coimbra:2011ky}
\begin{equation}
 [\nabla_M,\nabla_P]V^P+
\frac{1}{2}\nabla_A(Y^A{}_M{}^B{}_P\nabla_BV^P)={\cal
R}_{MR}V^R\, .
\end{equation}
Namely, the generalized Ricci tensor can be expressed as a commutator of
covariant derivatives plus a term proportional to the invariant $Y$ that,
as pointed out above, measures  in some sense the departure from ordinary
Riemannian
geometry. Interestingly enough, due to the section condition  the operator
$\nabla_A(Y^A{}_M{}^B{}_P\nabla_B \cdot )$  has
no second order derivatives. Moreover, when the Ricci is projected to the space of deformations of the generalized metric, its undetermined pieces get projected out \cite{Coimbra:2011ky}.

\subsection{Generalized Ricci tensor and twisted conditions} \label{sec:twistedgenRicci}
In this section we  assume that all vectors take
the form (\ref{vectors}),
 and consider diffeomorphisms that close under the twisted constraints
of Section  \ref{twistedconstraints}.

The starting point is the failure of the
Christoffel connection to transform as a tensor
\be
\Delta_{\xi} \Gamma_{MP}{}^Q = A^Q{}_P{}^R{}_S\ \partial_M \partial_R
\xi^S - \frac 1 2 \delta^Q_P  \partial_M \partial_R \xi^R- Y^N{}_M{}^R{}_S
\ \Omega_{R L}{}^S \Omega_{NP}{}^Q\ \xi^L \ .
\ee
It can be verified that its trace (\ref{trace}) transforms properly,
provided the twisted constraints hold
\be
\Delta_{\xi} \Gamma_{NM}{}^N - \Delta_\xi (-2 \partial_M \Delta) = -\frac
1 {28}\ \xi^P \Delta_{PMN}{}^N = 0\ .
\ee

In this case, for the Riemann tensor we get
\bea
\Delta_{\xi} R_{MNK}{}^L &=& \left(2 \Delta_{E_{\bar A}} \Gamma_{[MN]}{}^Q
\Gamma_{QK}{}^L \right.\label{FailureRiemann}\\
& &- 2\ Y^T{}_{[M|}{}^R{}_S \ \partial_R E_{\bar A}{}^S\
\partial_T\Gamma_{|N]K}{}^L   - 2 \ Y^T{}_{[M|}{}^R{}_S \partial_{|N]}
(\Omega_{R\bar A}{}^S \Omega_{TK}{}^L)\nn\\
&& \left.- 2\ Y^T{}_{[M|}{}^R{}_S\ \Omega_{R\bar A}{}^S
\left(\Omega_{TK}{}^O
\Gamma_{|N]O}{}^L - \Omega_{T|N]}{}^O \Gamma_{OK}{}^L - \Omega_{TO}{}^L
\Gamma_{|N]K}{}^O\right)\right) \xi^{\bar A}\ .\nn
\eea
So again we find that the usual Riemann tensor (i.e. in ordinary general
relativity) is not covariant under the diffeomorphisms that close under the twisted constraints.
The same happens for the usual Ricci tensor, because tracing the above
expression does not solve the problem
\be
R_{MN} = R_{MPN}{}^P \ , \ \ \ \ \ \ \Delta R_{MN} = \Delta R_{MPN}{}^P
\neq 0\ .
\ee
However, the Ricci tensor can be generalized into a (symmetric)
generalized
Ricci tensor by slightly extending the definition
(\ref{generalizedRicciTensor})
 \be
{\cal R}_{MN} \equiv \frac 1 2 \left( R_{MN} + R_{NM} + \Gamma_{R M}{}^P\
Y^R{}_P{}^S{}_Q \ \Gamma_{SN}{}^Q - \Omega_{R M}{}^P\ Y^R{}_P{}^S{}_Q \
\Omega_{SN}{}^Q\right) = {\cal R}_{NM} \ .\label{generalizedRicciTensor2}
\ee
The last term vanishes if the section condition is imposed, which is not
the case here. This term must therefore be added to define a covariant Ricci tensor.
The trace of
this
term is the analogue of the term added in DFT in
\cite{Grana:2012rr}.

To check the covariance of the generalized Ricci tensor up to twisted
constraints, it is instructive to use  planar indices. We first write
the
Riemann tensor in planar indices
\be
R_{\bar C\bar D\bar A}{}^{\bar B} = 2 \partial_{[\bar C} W_{\bar D]\bar
A}{}^{\bar B} - 2 \Omega_{[\bar C \bar D]}{}^{\bar E} W_{\bar E \bar
A}{}^{\bar
B} - 2 W_{[\bar C|\bar A}{}^{\bar E} W_{|\bar D] \bar E}{}^{\bar B}\ ,
\ee
in terms of which the generalized Ricci tensor
(\ref{generalizedRicciTensor2})
takes the form
\be
2\, {\cal R}_{\bar A \bar B} = R_{\bar A \bar D \bar B}{}^{\bar D} +
R_{\bar B
\bar D \bar A}{}^{\bar D} + (W - \Omega)_{\bar D \bar A}{}^{\bar E}
Y^{\bar
D}{}_{\bar E}{}^{\bar F}{}_{\bar G} (W - \Omega)_{\bar F\bar B}{}^{\bar G}
-
\Omega_{\bar D \bar A}{}^{\bar E} Y^{\bar D}{}_{\bar E}{}^{\bar F}{}_{\bar
G}
\Omega_{\bar F\bar B}{}^{\bar G} \ .
\ee
Using the following identity
\be
Y^{\bar D}{}_{\bar E}{}^{\bar F}{}_{\bar G}\ \Omega_{\bar F\bar B}{}^{\bar
G} =
F_{\bar B \bar E}{}^{\bar D} - 2 \Omega_{[\bar B \bar E]}{}^{\bar D}\ ,
\ee
the generalized Ricci tensor can be recast in the form
\bea
2\, {\cal R}_{\bar A \bar B} &=& 2 W_{\bar E (\bar A}{}^{\bar D} (W -
F)_{\bar B)
\bar D}{}^{\bar E}  +
W_{\bar D \bar A}{}^{\bar E} Y^{\bar D}{}_{\bar E}{}^{\bar F}{}_{\bar G}
W_{\bar
F \bar B}{}^{\bar G} - 2 W_{\bar D \bar E}{}^{\bar D} W_{(\bar A \bar
B)}{}^{\bar E}\nn\\&&
- 2 \partial_{\bar D} W_{(\bar A \bar B)}{}^{\bar D} + 2 \partial_{(\bar A
|} W_{\bar D | \bar B)}{}^{\bar D}\ .\label{RicciTensorPlano}
\eea
Here, the first line is manifestly covariant, because  (the
planar version of) both the
spin connection and the fluxes are covariant, up to the twisted constraints. The
covariance in the second line is less trivial, because derivatives of
scalars are only tensors when the section condition holds
\be
\Delta_\xi (\partial_M \phi) = Y^R{}_M{}^P{}_Q \partial_P \xi^Q \partial_R
\phi\ ,
\ee
which is not an assumption in this section. Notice however, that the last
two terms can be re-written in an explicitly covariant form
\be
- 2 \partial_{\bar D} W_{(\bar A \bar B)}{}^{\bar D} + 2 \partial_{(\bar A
|} W_{\bar D | \bar B)}{}^{\bar D} = 2 W_{\bar Q \bar N}{}^{\bar Q}
W_{(\bar A \bar B)}{}^{\bar N} -8 \vartheta_{\bar A} \vartheta_{\bar B} -2
 \nabla_N V^N{}_{(\bar A \bar B)}\ ,
\ee
where $V^N{}_{(\bar A \bar B)}$ is a tensor defined by
\be
V^N{}_{(\bar A \bar B)} =  E_{\bar C}{}^N W_{(\bar A \bar B)}{}^{\bar C} +
2 E_{(\bar A}{}^N \vartheta_{\bar B)}\ .
\ee

We finish this section by giving an explicit expression for a covariant
generalized Ricci tensor
\bea
2\, {\cal R}_{\bar A \bar B} &=& 2 W_{\bar E (\bar A}{}^{\bar D} (W -
F)_{\bar B)
\bar D}{}^{\bar E}  +
W_{\bar D \bar A}{}^{\bar E} Y^{\bar D}{}_{\bar E}{}^{\bar F}{}_{\bar G}
W_{\bar
F \bar B}{}^{\bar G} -8 \vartheta_{\bar A} \vartheta_{\bar B} -2  \nabla_N
V^N{}_{(\bar A \bar B)}\ .\ \ \ \ \label{RicciTensorPlano3}
\eea
When tracing this expression and integrating with the measure to obtain an
action, the last term gives a total derivative and therefore vanishes. We
emphasize that this generalized Ricci tensor was constructed only
imposing the twisted constraints, and reduces to
(\ref{generalizedRicciTensor}) if the section condition is imposed.

\subsection{Generalized Ricci scalar and scalar potential of maximal
supergravity} \label{sec:RicciscalarN-8}

As we show in the Appendix \ref{AppD}, when taking the trace of the
generalized
Ricci tensor (\ref{generalizedRicciTensor2}) with the generalized  metric
$H$
\be
{\cal R} = {H}^{\bar A \bar B} {\cal R}_{\bar A \bar B} \ ,
\ee
the undetermined pieces of the connections drop out, and it can be
expressed
purely in terms of fluxes (no constraints are imposed in this derivation).
In the particular
case of $\vartheta_A = 0$,
we find
\be
\frac 1 4 {\cal R} = \frac 1{672} \left({H}^{AD}{H}^{BE}{H}_{CF}
X_{AB}{}^C
X_{DE}{}^F + 7 {H}^{AB} X_{AC}{}^D X_{BD}{}^C\right) \ .\label{scalarpot}
\ee
Remarkably, this takes the exact same form as the scalar potential of
gauged
maximal supergravity \cite{de Wit:2007mt} if we identify the generalized
metric
with the moduli space matrix ${\cal M}$. Note also that this is true for
any
torsionless and metric-compatible connection, and the concrete expression
of the
determined part does not need to be known. In fact, we never needed to
solve for
the spin connection, but only used the equation that defines it
implicitly.

Finally, notice that by definition the Ricci scalar transforms indeed as a
scalar under generalized diffeomorphisms
\be
\delta_{\xi} {\cal R} = \xi^P \partial_P {\cal R}\ .
\ee
This can also be checked taking into
account that the fluxes are covariant provided the quadratic constraints
hold.
Combining this with the fact that $e^{-2\Delta}$ transforms as a density
(\ref{measure})
\be
\delta_\xi e^{-2\Delta} = \partial_P (e^{-2\Delta} \xi^P)\ ,
\ee
we arrive at the action of EFT
\be
S = \frac 1 4 \int d^{56}Y\ e^{-2\Delta}\ {\cal R}\label{actionfinal} ,
\ee
which is invariant under generalized diffeomorphisms
(\ref{InternalGenDiffs}). In Appendix \ref{AppD} we provide a detailed
derivation of (\ref{scalarpot}). Form  (\ref{actionfinal}) we can see that
in the context of string theory or M theory, when the section condition holds,
 $e^{-2\Delta}$ can be identified with the
measure of the internal $6$ or $7$-dimensional manifold
\be
e^{-2\Delta}\propto \sqrt{g} \ .
\ee

\section{Summary and outlook}
\label{sec:conlusions}

In this work we explored the U-duality covariant framework of extended geometry, focusing on
the case of $E_7$, and applied it to describe
the moduli space of maximal gauged supergravity in
four dimensions. The extended space is a $\bf
56$-dimensional mega-space equipped with a generalized bein taking values
in $\mathbb{R}^+\times E_7/SU(8)$, which can be parameterized in terms of
Type II or M-theory degrees of freedom. The first step in the
construction is the introduction of generalized gauge transformations
(or generalized diffeomorphisms)
(\ref{InternalGenDiffs}), which unify all the
possible gauge transformations of the theory inherited from the metric, NSNS and RR
forms in Type II strings or from the 3-form in M-theory.

When the generalized diffeomorphisms act on the bein, one
obtains field-dependent (i.e. non constant) fluxes (\ref{fluxesplanar}), which are in the
$\bf 56 + \bf 912$ irreducible representations of $E_7$. For consistency, the generalized
diffeomorphisms must satisfy a set of conditions, such as closure of the gauge algebra
(\ref{Closure}). We showed that, as happens in DFT, these constrains
allow for at least two different type of ``solutions''. One of them are
configurations obeying the section condition (\ref{sectioncondition}), which
implies that the fields only depend on coordinates spanning a 7-dimensional slice of the extended
space, therefore allowing to make contact with supergravity and
Exceptional Generalized Geometry. In the other type of solutions, which
we call ``twisted'', the fields are taken to have a Scherk-Schwarz form,
and the constraints translate into constraints for the fluxes.
Interestingly, in the case of constant fluxes they match the quadratic
constraints of maximal supergravity, but more generally we provide the
extension to the case of non-constant fluxes. The advantage of this
second approach is that it allows for truly extended configurations, with
dependence on the extra coordinates. Duality
orbits of gaugings allowed in maximal gauged supergravity which are beyond those
coming from conventional compactifications (and their dual configurations) can be reached
in this way.

We then described the geometry of the extended space, starting from a derivative transforming
covariantly under generalized diffeomorphisms, and
their corresponding Christoffel and spin connections. These are not uniquely defined, only a
subset of its components are, via the torsionless and metric compatibility conditions. We
summarized the properties of the different connections in Table 1.
The next question is whether a curvature for the extended space
can be defined. Since there seems to be no easy way to define a covariant
generalized Riemann tensor \cite{Coimbra:2011nw}, we considered only the Ricci tensor and
scalar, which transform appropriately up to the constraints of the theory.
We showed how to meet these conditions in two different ways. In the first one, the Ricci tensor
 behaves appropriately if the section condition is imposed, and it corresponds to
the natural generalization of that introduced in the context of Double
Field Theory \cite{Jeon:2011cn,Hohm:2011si} and equals that of \cite{Coimbra:2011ky} in the context of generalized ($\mathbb{R}^+\times E_{7(7)}$) geometry. In the second approach, the section condition is
relaxed within
the context of Scherk-Schwarz-like compactifications, and we showed that
the definition of the Ricci tensor must be further extended so that it is also covariant up to
the twisted constraints (\ref{Delta5}). Up to our knowledge, this is the first covariant
construction of an extended geometry where the section condition is not
imposed.

Finally, we showed that the resulting Ricci scalar matches exactly the scalar
potential of gauged maximal supergravity, provided
one associates the generalized metric with the moduli scalar matrix,
the dynamical fluxes are taken to be the constant gaugings, and the gaugings in the $\bf 56$ are
taken to zero. Although the
original expression for the generalized Ricci scalar
(\ref{generalizedRicciTensor2}) is a function of the spin connection,
which contains undetermined pieces, we show that these contributions
simply drop out, and therefore any torsionless and metric compatible
connection gives the desired Ricci scalar, whatever its undetermined part is.

Let us finally comment on some interesting questions that remain open, and
are worth exploring in our point of view. It is known that the
section condition implies that this framework is a covariant re-writing of
higher-dimensional supergravity compactifications. In this kind of
compactifications only a subset of the gaugings can be reached in the lower
dimensional effective action, and thus the space of gauging orbits is
split into those that can be obtained (geometric) and those that cannot
(non-geometric) \cite{Dibitetto:2012rk}. Restricting to Scherk-Schwarz-type backgrounds,
this construction does not
necessarily use the section condition, and instead uses the twisted
conditions, which are in one to one correspondence with the constraints of
gauged supergravity. Therefore, any orbit of gaugings can be reached
geometrically in this construction, even those that are non-geometric from
a supergravity point of view. Recently there has been much progress in
moduli-fixing, fluxed induced supersymmetry breaking, de Sitter vacua
surveys, etc. in the
presence of non-geometric gaugings, and we believe that this framework can
shed light on the higher dimensional uplift of these orbits, as background
fluxes on the mega-space. For example, recently a one-parameter family of
new maximal gauged supergravity with $SO(8)$ gauge group was found
\cite{Dall'Agata:2012bb}. It would be nice to seek for an uplift of these gaugings to the
mega-space considered here.

{\bf Note added}: After our work appeared on the ArXiv the preprint \cite{cederwall},
with substantial overlap with our sections 4 and 5, was uploaded.

\section*{Acknowledgments}
{We are very grateful to C. Nu\~{n}ez for collaboration in the early
stages
of this project. We also thank E. Andr\'es, G. Dibitetto, V. Penas, M. Trigiante  and D.
Waldram  for
useful discussions and
comments. G.A. thanks the ICTP for hospitality during the
completion of this work. This work was partially supported
by CONICET,  the ERC Starting Grant 259133 -- ObservableString and
EPLANET.}

\appendix

\section{Useful $E_7$ identities}\label{AppA}

The $e_{7(7)}$ algebra with
generators $(t_\alpha)_M{}^N$  where $\alpha = 1,\dots,133$ is an index
in the
adjoint $\bf 133$, and $M,N,\dots = 1,\dots,56$. The indices are raised
and
lowered with the symplectic $Sp(56)\supset E_7$ metric $\omega_{MN}$
according
to the conventions (\ref{convention}). With this in mind, the adjoint of
$E_7$
is symmetric $(t_\alpha)_{(MN)}$.

The symplectic metric $\omega_{MN}$ is left invariant by $E_7$
transformations,
as is the quartic invariant $K_{MNPQ}$. Contracting two generators, we can
define a projector to the adjoint
representation
\be
P_{(adj)MNPQ} = (t_\alpha)_{MN}(t^\alpha)_{PQ}= \frac{1}{12} \omega_{M(P}
\omega_{Q)N} + K_{MNPQ}\  ,\label{projector133}
\ee
satisfying the useful identities
\be
P_{(adj)}{}^M{}_N{}^P{}_Q \ P_{(adj)}{}^Q{}_P{}^R{}_S =
P_{(adj)}{}^M{}_N{}^R{}_S \ , \ \ \ \ P_{(adj)}{}^M{}_N{}^N{}_M = 133
\label{PP=P}
\ee
\begin{equation}
P_{(adj)MN}{}^{PQ} = P_{(adj)(MN)}{}^{(PQ)} = P_{(adj)}{}^{PQ}{}_{MN}\ ,
\end{equation}
and
\be
P_{(adj)M}{}^K{}_N{}^L = \frac 1 {24} \delta^K_M \delta^L_N + \frac 1 {12}
\delta^L_M \delta^K_N - \frac{1}{24} \omega_{MN} \omega^{KL} +
P_{(adj)MN}{}^{KL} \ . \label{idPadj}
\ee
Also, using (\ref{PP=P}) one can show
\be
K_{MN}{}^{PQ} K^{MNKL} = - \frac 5 6 K^{PQKL} - \frac {11}{ 12 \times 12}
\omega^{P(K} \omega^{L)Q} \ .\label{KK=K}
\ee
A very useful identity to show the relation between the relaxed
constraints and the section condition is
\be
12 P_{(adj)}{}^{(MN}{}_{QT} P_{(adj)}{}^{P)T}{}_{RS} - 4 K^{MNPT}
P_{(adj)TQRS}
+ P_{(adj)}{}^{(MN}{}_{RS} \delta^{P)}_Q = 0\ ,\label{CyclicPadj}
\ee
and the final useful properties we used in the paper are
\be
(t_\beta)_M{}^Q (t^\alpha)_Q{}^P (t^\beta)_P{}^N = \frac 7 8
(t^\alpha)_M{}^N \ , \ \ \ \ \ \ (t_\alpha)_L{}^{(P} K^{QRS)L} = 0\ .
\label{propsE7}
\ee
\subsection{$SU(8)$ subgroup}

The maximal compact subgroup of $E_7$ is  $SU(8)$. When $E_7$ is broken to
$SU(8)$, the fundamental $\bf 56$ and adjoint $\bf 133$ representations
break
according to
\be
M = {\bf 56} = {\bf 28} + \overline{\bf 28} = (_{[ij]},^{[ij]})\ , \ \ \ \
\ \
\alpha = {\bf 133} = {\bf 63} + {\bf 70} = (_i{}^j,_{[ijkl]})\ ,
\ee
where $i,j,\dots = 1,\dots,8$ and the $\bf 63$ is traceless $_i{}^i = 0$.

The 133 generators of $E_7$  break into 63 and 70 generators, respectively
\cite{LeDiffon:2008sh}
\bea
(t_i{}^j)_{mn}{}^{kl} &=& - \delta^j_{[m} \delta^{kl}_{n]i} - \frac 18
\delta^j_i \delta^{kl}_{mn} = - (t_i{}^j)^{kl}{}_{mn}\nn\\
(t_{ijkl})_{mnpq} &=& \frac 1{24} \epsilon_{ijklmnpq} \ , \ \ \
(t_{ijkl})^{mnpq} = \delta^{mnpq}_{ijkl} \ ,\label{generatorsE7su8}
\eea
with Cartan-Killing metric
\be
\kappa_m{}^n,_p{}^q = 3 (\delta^q_m \delta^n_p - {\frac 18} \delta^n_m
\delta^q_p) \ , \ \ \ \ \ \kappa_{ijkl},_{mnpq} = \frac 1 {24}
\epsilon_{ijklmnpq} \label{Killing} \ .
\ee

The projection to the adjoint in the product ${\bf 56} \times {\bf 56}$
reads
\begin{align} \label{56x56=133SLE}
(V\cdot \hat V)_{i}{}^{j}&=(V^{kj} {\hat
V}_{ki}-\frac{1}{8}\delta^{j}_{i}V^{kl}
{\hat V}_{kl})+(\hat V^{kj} V_{ki}-\frac{1}{8}\delta^{j}_{i}\hat V^{kl}
V_{kl})\\
(V\cdot \hat V)_{ijkl}&=-3( V_{[ij} {\hat V}_{kl]} + \frac{1}{4!}
\epsilon_{ijklmnop} V^{mn} \hat V^{op}) \nn \ .
\end{align}

\subsection{$SL(2) \times O(6,6)$ subgroup} \label{AppA:SL2O66}

The fundamental ${\bf 56}$ representation of $E_7$ splits according to its
$SL(2) \times O(6,6)$ subgroup as follows
\bea
{\bf 56}&=& ({\bf 2}, {\bf 12}) \oplus ({\bf 1},{\bf 32}) \nn \\
M &=& \ (\hat \imath m) \ \oplus \ \  \alpha\ ,
\eea
where $\hat \imath =+,-$ is a fundamental $SL(2)$ index, $m$ an $O(6,6)$
index,
and $\alpha$ is an $O(6,6)$ Majorana-Weyl spinor index.

The symplectic metric decomposes as
\be
\Omega_{MN} = \left(\begin{matrix} \epsilon_{\hat \imath \hat \jmath }
\eta_{mn}& \\ & C_{\alpha \beta}\end{matrix}\right) \ , \ \ \
\epsilon_{\hat
\imath \hat \jmath } = \left(\begin{matrix} 0& -1\\ 1&
0\end{matrix}\right)\ ,
\ \ \ \eta_{mn} = \left(\begin{matrix} 0& 1\\ 1&  0\end{matrix}\right)\ .
\ee

The different components of the projector onto the ${\bf 133}$
representation
read \cite{halving}
\beq
\begin{array}{cclc}
P_{(adj)}{}_{\hat \imath m \hat \jmath n}{}^{\hat k p \hat l q} & = &  -
\tfrac{1}{2} \, \epsilon_{\hat \imath \hat \jmath} \, \epsilon^{\hat k
\hat l}
\, \delta_{mn}^{pq} + \tfrac{1}{12} \, \delta_{\hat \imath}^{(\hat k} \,
\delta_{\hat \jmath}^{\hat l)} \, \eta_{mn}\, \eta^{pq}&  \\[4mm]
P_{(adj)}{}_{\hat \imath m \hat \jmath n}{}^{\alpha \beta} & = & -
\tfrac{1}{8}
\, \eps_{\hat \imath \hat \jmath} \, \left[ \gamma_{mn} \right]^{\alpha
\beta} &
 \\[4mm]
P_{(adj)}{}_{\alpha \beta}{}^{\gamma \delta} & = & - \tfrac{1}{32} \,
\left[
\gamma_{mn} \right]_{\alpha \beta} \, \, \left[ \gamma^{mn}
\right]^{\gamma
\delta} &  \\[4mm]
P_{(adj)}{}_{\hat \imath m \alpha}{}^{\hat \jmath n \beta} & = &
\tfrac{1}{24}
\, \delta_{\hat \imath}^{\hat \jmath} \, \left( \, {\left[
{\gamma_{m}}^{n}
\right]_{\alpha}}^{\beta}  +  {\delta_{m}}^{n} \,
{\delta_{\alpha}}^{\beta}
\right)& .
\end{array} \label{adjprojTsub}
\eeq

\section{The Weitzenb\"ock connection and the algebra}\label{AppB}
Given an element $E$ of $E_7$
\be
E = \exp(\phi^\alpha t_\alpha)\ ,
\ee
where $t_\alpha$ are the generators of $G$
\be
[t_\alpha , t_\beta] = f_{\alpha \beta}{}^\gamma t_\gamma\ ,
\ee
the Weitzenb\"ock connection, defined as
\be
\Omega_{ M  N}{}^{ P} = - \partial_{ M}(E^{-1})_{N}{}^Q\ E_Q{}^{P}\ ,
 \ee
is an element of the algebra of $G$. This can be easily seen by use of the
identity
\be
\partial_M e^X . e^{-X} = \partial_M X + \frac{1}{2 !} [X, \partial_M X] +
\frac{1}{3 !} [X, [X, \partial_M X]] + \dots\ .
\ee
A quick computation shows that
\be
\Omega_{ M   N}{}^{  P} = \Omega_{  M}{}^\alpha (t_{\alpha})_N{}^P\ ,
\label{WizenAlgebra}
\ee
with
\be
\Omega_{   M}{}^\alpha = \partial_{  M} \phi^\alpha - \frac 1{2!}
f_{\sigma\beta}{}^\alpha \phi^\sigma \partial_{  M} \phi^\beta + \frac 1
{3!}
f_{\mu\rho}{}^\sigma f_{\beta \gamma}{}^\rho \phi^\mu \phi^\beta
\partial_{  M}
\phi^\gamma - \dots
\ee
\section{Representations, projectors and generalized
diffeomorphisms}\label{AppC}

In this Appendix we first present the projectors onto the
irreducible representations in the tensor product of the fundamental
with the adjoint representation of an arbitrary simple group G, following the
appendix of \cite{deWit:2002vt}. Related expressions and useful identities
can
be found in \cite{Riccioni:2007au}. We will then show how this sheds light
in
the interpretation of the coefficients appearing in the structure of
generalized
diffeomorphisms, in terms of group theoretical quantities.

\subsection{Representations and Projectors}

For any simple group (with the exception of $E_8$),  the product of a
fundamental
representation ${\bf
D(\Lambda)}$ times the adjoint decomposes in the direct sum of ${\bf
D(\Lambda)}$ plus two other representations, ${\bf D_1}$ and~${\bf
D_2}$, with ${\rm dim}({\rm D}_1)< {\rm dim}({\rm D}_2)$,
\begin{eqnarray}
{\bf D(\Lambda)}\times {\bf Adj(G)}\rightarrow
{\bf D(\Lambda)}+{\bf D_1}+{\bf D_2}  \ .
\label{gendec}
\end{eqnarray}
 This is also true for
orthogonal groups by replacing  the fundamental representation by
the spinor representation.  Supersymmetry requires
\beq
F_{MN}{}^P \in {\bf D(\Lambda)}+{\bf D_1}\ ,
\eeq
and therefore it is useful to construct projectors onto these
representations.

Let us call  $d_\Lambda={\rm dim}({\bf
D(\Lambda)})$, $d={\rm dim}({\rm G})$, and $\{t^\alpha\}$
($\alpha=1,\dots , d$) the generators of G in the ${\bf D(\Lambda)}$
representation. Furthermore, let $C_\theta,\,C_\Lambda$ be the
Casimirs of the adjoint and fundamental representations,
respectively. The invariant matrix $\eta^{\alpha\beta}={\rm
Tr}(t^\alpha t^\beta)$ is used to rise and lower the adjoint
indices, and  is related to the Cartan-Killing metric
$\kappa^{\alpha\beta}$ by
\begin{eqnarray}
\kappa^{\alpha\beta}&=&
\frac{d}{C_\Lambda d_\Lambda}\, \eta^{\alpha\beta}\ .
\end{eqnarray}
Using the definition of the Casimir operator,
$C_\Lambda\,1_{d_\Lambda} =\kappa_{\alpha\beta} t^\alpha
t^\beta$, we have the following relation
\begin{eqnarray}
f_{\alpha\beta}{}^\gamma\, f^{\alpha\beta}{}_\sigma
&=&-\frac{d}{d_\Lambda}\, C_{\rm r}\,\delta_{\sigma}^\gamma
\;,\qquad
\mbox{with} \quad
C_{\rm r}=\frac{C_\theta}{C_\Lambda}=
\frac{d_\Lambda}{d}\frac{g^{\vee}}{\tilde{I}_\Lambda}\ ,
\;,
\end{eqnarray}
where $g^{\vee}$ is the dual Coxeter number and $\tilde{I}_\Lambda$
is the Dynkin index of the fundamental representation.  In the simply
laced case we have additionally
\begin{eqnarray}
C_{\rm r}&=&\frac{d_\Lambda}{d}\left(
\frac{d}{r}-1\right)\frac{1}{\tilde{I}_\Lambda}
\ ,
\end{eqnarray}
with $r$ the rank of G.

Denote the projectors on the representations in (\ref{gendec}) by
$P_{({D(\Lambda)})},\,P_{({D_1}) },\,
P_{({D_2}) }$. These are orthonormal, i.e.
\be
P_{(X)M}{}^{\alpha P}{}_\gamma \ P_{(Y)P}{}^{\gamma N}{}_\beta =
\delta_{XY} \
P_{(X)M}{}^{\alpha N}{}_\beta\ ,
\ee
and sum to the identity on ${\bf
D(\Lambda)}\times {\bf Adj(G)}$.  These three projectors can be
expressed in terms of three independent objects, namely:
\begin{eqnarray}
P_{({D(\Lambda))M}}{}^{\alpha N} {}_\beta
&=&\frac{d_\Lambda}{d}\, (t^\alpha t_\beta)_M{}^N\,,\nonumber\\
P_{({D_1})M }{}^{\alpha N} {}_\beta &=& a_1 \,\delta^\alpha
{}_{\!\beta}\,\delta_M{}^N+a_2 \, (t_\beta t^\alpha)_M{}^N+a_3\,
(t^\alpha
t_\beta)_M{}^N\,,\nonumber\\
P_{({D_2})M }{}^{\alpha N} {}_\beta&=&
(1-a_1)\,\delta^\alpha{}_{\!\beta}\, \delta_M{}^N-a_2\, (t_\beta
t^\alpha)_M{}^N-({d_\Lambda/d}+a_3)\,(t^\alpha t_\beta)_M{}^N
\ ,
\label{projectors}
\end{eqnarray}
where
\begin{eqnarray}
a_1 &=&\frac{d_\Lambda\left(4+(C_{\rm r}-4)
d)\right)+d_1\left((C_{\rm r}-2)
d-2)\right)}{\left(10+d(C_{\rm r}-8)+d^2 (C_{\rm r}-2)\right)
d_\Lambda}\,, \nonumber\\
a_2 &=&-\frac{2\left(4+(C_{\rm r}-4)
d)\right)\left((d-1)d_\Lambda-2d_1\right)}{\left(10+d(C_{\rm
r}-8)+d^2 (C_{\rm r}-2)\right) C_{\rm r} d }\,, \nonumber\\
a_3 &=&\frac{-d_\Lambda\left(4+(C_{\rm r}-4)
d)\right)\left(2+(C_{\rm r}-2)d\right)+d_1\left(16(d-1)-10 (d-1)
C_{\rm r}+C_{\rm r}^2 d \right)}{\left(10+d(C_{\rm r}-8)+d^2 (C_{\rm
r}-2)\right) C_{\rm r} d } \ ,
\nonumber
\end{eqnarray}
with $d_1=\dim({\bf D_1})$. Moreover, $d_1$ is determined to be
\begin{eqnarray}
d_1 &=& \frac{d_\Lambda}{2}\left[d-1+\frac{\sqrt{C_{\rm
r}}\left(10+d(C_{\rm r}-8)+d^2 (C_{\rm r}-2)\right) }{\sqrt{256
(d-1)+C_{\rm r} (100+4 d (5 C_{\rm r}-38)+(C_{\rm r}-2)^2
d^2)}}\right] \ .{\quad{~}}
\label{Delta}
\end{eqnarray}
In table \ref{projs}, taken from \cite{ deWit:2002vt}, we give these
coefficients for all simple
Lie algebras except ${\rm E}_{8}$ (for which the relevant projectors
have been computed in \cite{Koepsell:1999uj}).

\begin{table}
\begin{center}
\begin{tabular}{l c c l l r r r}\hline
G & $g^{\vee}$ & $d_\Lambda$ & $\tilde{I}_\Lambda $ & $d_1$       &
$a_1$   & $a_2$ & $a_3$ \\
\hline
${\rm A_r}$ & $r+1$ & $r+1$ &$\frac{1}{2}$    &
$\frac{1}{2}(r-1)(r+1)(r+2)$ &  $\frac{1}{2}$ &  $-\frac{1}{2}$&
$-\frac{1}{2 r}$  \\
${\rm B_r}$   & $2r-1$ & $2r+1$& 1    & $\frac{1}{3}r (4 r^2-1)$  &
$\frac{1}{3}$ & $-\frac{2}{3}$ & $0$ \\
${\rm B_r}$   & $2r-1$ & $2^{r}$& $2^{r-3}$    & $2^{r+1}\,r$  &
$\frac{2}{2r-1}$ & $-2^{r-1}\frac{1}{2r-1}$ & $2^{r-1}\,\frac{2r-7}{4
r^2-1}$ \\
 ${\rm C_r}$   & $r+1$ & $2r$& $\frac{1}{2}$    & $\frac{8}{3}r
(r^2-1)$  & $\frac{2}{3}$ & $-\frac{2}{3}$ & $-\frac{2}{1+2r}$ \\
${\rm D_r}$   & $2r-2$ & $2r$ & 1    & $\frac{2}{3}r (2r^2-3 r +1)$  &
$\frac{1}{3}$ &  $-\frac{2}{3}$ & $0$ \\
${\rm D_r}$   & $2r-2$ & $2^{r-1}$ &  $2^{r-4}$   & $2^{r-1}\,(2r-1)$
& $\frac{1}{r-1}$ &  $-2^{r-3}\frac{1}{r-1}$ &
$2^{r-3}\frac{(r-4)}{r\,(r-1)}$ \\
 ${\rm G_2}$   & 4 & 7 & 1    & $27$  & $\frac{3}{7}$ &
$-\frac{6}{7}$ & $-\frac{3}{14}$ \\
 ${\rm F_4}$   & 9 & 26 & 3    & $273$  & $\frac{1}{4}$ &
$-\frac{3}{2}$ & $\frac{1}{4}$ \\
${\rm E}_{6}$ & 12 & 27 & 3 & 351  &  $\frac{1}{5}$  & $-\frac{6}{5}$
& $\frac{3}{10}$ \\
${\rm E}_{7}$ & 18 & 56 &6       & 912   & $\frac{1}{7}$ &
$-\frac{12}{7}$ &   $\frac{4}{7}$  \\
\hline
\end{tabular}
\end{center}
\caption{\small
Coefficients needed to construct the projectors for
all simple algebras except $E_8$. }
\label{projs}
\end{table}

\subsection{Generalized diffeomorphisms} \label{Appalphacoeff}

Let us show how these results shed light on the general structure of
generalized
diffeomorphisms, equation (\ref{InternalGenDiffs}), in particular why is the
proportionality coefficient between the tensor $A$ and the projector to
the
adjoint representation equal to 12 for $E_7$ (see equation (\ref{A})), and what
it
would be for other groups.
Let us write the generalized diffeomorphisms (\ref{InternalGenDiffs}) in
the
generic form, as in \cite{Berman:2012vc}
\be
{\cal L}_\xi V^M = \xi^P \partial_P V^M - \alpha P_{(adj)}{}^M{}_N{}^P{}_Q
\partial_P \xi^Q V^N + \frac \omega 2  \partial_P \xi^P V^M\ ,
\ee
where $\alpha$ is some group-dependent constant, and the projector to the
adjoint is given in (\ref{A}).
Restricted to orthogonal frames $E_{\bar A}$, these transformations must
reproduce the embedding tensor components that are compatible with
supersymmetry, i.e., the ${\bf D(\Lambda)}+{\bf D_1}$ components, and
project
out the remaining representation ${\bf D_2}$. Let us set $\omega = 0$ for
the moment, i.e., let us assume that the global symmetry group has no
$\mathbb{R}^+$ component (later we will restore $\omega$). In this case we
find
\be
F_{MN}{}^P  = E^{\bar A}{}_ M E^{\bar B}{}_N {\cal L}_{E_{\bar A}} E_{\bar
B}{}^P = \Omega_{MN}{}^P - \alpha P_{(adj)}{}^P{}_N{}^K{}_L
\Omega_{KM}{}^L\ .
\ee
As we showed in (\ref{AppB}), the Weitzenb\"ock connection $\Omega$ takes
values in the fundamental times the adjoint of the global symmetry group,
so
\be
F_{MN}{}^P = F_{M}{}^{\alpha} (t_{\alpha})_N{}^P\ ,
\ee
with
\be
F_M{}^\alpha = Q_M{}^\alpha,^N{}_\beta \Omega_N{}^\beta \ , \ \ \ \ \ \
Q_M{}^\alpha,^N{}_\beta = \delta^N_M \delta^\alpha_\beta - \alpha (t_\beta
t^\alpha)_M{}^N\ .
\ee
The coefficient $\alpha$ must then be fixed in such a way that the tensor
$Q_M{}^\alpha,^N{}_\beta$ is a linear combination of the projectors to
${\bf
D(\Lambda)}+{\bf D_1}$. We find
\be
Q_M{}^\alpha,^N{}_\beta = \frac 1 {a_1}  P_{{(D_1)}M }{}^{\alpha N}
{}_\beta -
\frac{a_3}{a_1} \frac{d}{d_\Lambda} P_{{(D(\Lambda))M}}{}^{\alpha N}
{}_\beta\ ,
\label{ProjectorA}\ee
provided
\be
\alpha = - \frac{a_2}{a_1} = \frac{2 d_\Lambda\left(4+(C_{\rm r}-4)
d)\right)\left((d-1)d_\Lambda-2\Delta\right)}{[d_\Lambda\left(4+(C_{\rm
r}-4)
d)\right)+d_1\left((C_{\rm r}-2)
d-2)\right)] C_r d }\ ,\label{alpha}
\ee
corresponding in particular to $\alpha = 12$ in $E_7$, as stated in
equation
(\ref{A}).

Let us now see how the coefficient $\omega$ can be fixed. A possibility is
to demand  that the intertwining
tensor (i.e. the symmetric part of the gauge group generators
$F_{(MN)}{}^P$) takes values in the algebra of the global symmetry group
without the $\mathbb{R}^+$, as explained in
\cite{LeDiffon:2008sh}. Here we work out the $E_7$ case, but the other
cases
follow analogously. For a generic value of $\omega$, we can compute the
general form of the symmetric part of the gauge group generators, which
reads
\be
F_{(MN)}{}^P = - \omega \vartheta_{(M} \delta_{N)}^P +
(t_\alpha)_{(M}{}^P\Theta_{N)}{}^\alpha + 8 \vartheta_Q
P_{(adj)}{}^Q{}_{(M}{}^P{}_{N)} \ .\label{Intertwining}
\ee
We can now use the relation (\ref{idPadj}), together with the fact that
the $\bf 912$ satisfies
\be
(t_\alpha)_{(M}{}^P\Theta_{N)}{}^\alpha = - \frac 1 2 (t_\alpha)_{MN}
\Theta^{P\alpha}\ ,
\ee
to show that (\ref{Intertwining}) can be written as
\be
F_{(MN)}{}^P = (1 - \omega) \vartheta_{(M} \delta_{N)}^P - \frac 1 2
(\Theta^{P\alpha} -16 \vartheta_Q (t^{\alpha})^{PQ})(t_\alpha)_{MN}\ .
\ee
Here the first term measures the failure of the intertwining
tensor to take values in the algebra of $E_7$, and in order to cancel it
we must take
\be
 \omega = 1\ .
\ee
This procedure can be repeated for any other group analogously.

In supersymmetric theories, the representation $\bf D_2$ is projected out
from
the embedding tensor through a linear constraint. Given that the
projectors are
normalized to add to unity, the projectors to $\bf D(\Lambda)$ and $\bf
D_1$
contain information about the projector to $\bf D_2$. This means that the
coefficient $\alpha$ carries information about supersymmetry. The linear
constraint is automatically engineered in the definition of the
generalized Lie
derivative through (\ref{ProjectorA}), which is therefore consistent with
(and
encodes information of) supersymmetry.

Let us conclude this section to see how the generalized diffeos of Double
Field Theory and usual Riemannian geometry arise as particular examples of
these generalized expressions. For DFT, with gauge group $O(d,d)$, the
generators and projector to the adjoint are given by
\be
(t_{[MN]})_P{}^Q = \eta_{P[M} \delta_{N]}^Q \ , \ \ \ \ \
P_{(adj)}{}^M{}_N{}^P{}_Q = \frac 1 2 (\delta_N^P \delta^M_Q - \eta^{MP}
\eta_{NQ})\ ,
\ee
and we have $\alpha = 2$, so
\be
{\cal L}_\xi V^M = \xi^P \partial_P V^M - 2 P_{(adj)}{}^M{}_N{}^P{}_Q
\partial_P \xi^Q V^N =  \xi^P \partial_P V^M + (\partial^M \xi_P -
\partial_P \xi^M) V^P\ ,
\ee
which matches the expression of \cite{Hull}. For usual Riemannian
geometry, we have the group $GL(d)$ with generators and projector to the
adjoint
\be
(t_{M}{}^N)_P{}^Q = \delta_M^Q \delta^N_P \ , \ \ \ \ \
P_{(adj)}{}^M{}_N{}^P{}_Q = \delta_M^Q \delta^N_P\ ,
\ee
and we have $\alpha = 1$, so
\be
{\cal L}_\xi V^M = \xi^P \partial_P V^M - P_{(adj)}{}^M{}_N{}^P{}_Q
\partial_P \xi^Q V^N =  \xi^P \partial_P V^M - \partial_P \xi^M  V^P\ ,
\ee
which is the usual Lie derivative.
\section{The scalar potential from extended geometry}\label{AppD}

Here we show how the trace of the generalized Ricci tensor
(\ref{RicciTensorPlano})
\bea
2\, {\cal R}_{\bar A \bar B} &=& 2 W_{\bar E (\bar A}{}^{\bar D} (W -
F)_{\bar B)
\bar D}{}^{\bar E}  +
W_{\bar D \bar A}{}^{\bar E} Y^{\bar D}{}_{\bar E}{}^{\bar F}{}_{\bar G}
W_{\bar
F \bar B}{}^{\bar G} -8 \vartheta_{\bar A} \vartheta_{\bar B} -2  \nabla_N
V^N{}_{(\bar A \bar B)}\ , \ \ \ \ \label{RicciTensorAppD}
\eea
introduced in Section \ref{genRiccis} gives  the scalar potential of
maximal supergravity (\ref{scalarpot}). To prove this we will use the
following assumptions:
\begin{itemize}
\item The spin connection is torsionless (\ref{spincon912})
\be
W_{AB}{}^C - 12 P_{(adj)}{}^C{}_B{}^D{}_E \ W_{DA}{}^E = F_{AB}{}^C +
\vartheta_A \delta_B^C
\label{TorsionlessAppD} \ .
\ee
From here it is also clear that it belongs to the $\bf 56 \times 133$
representation $W_{AB}{}^C = W_A{}^\alpha (t_{\alpha})_B{}^C$. This also
implies that the trace is given by
\be
W_{BA}{}^B = - 2 \vartheta_A\ .\label{SpinTraceAppD}
\ee

\item The spin connection is generalized metric compatible
(\ref{metriccomp})
\be
W_{AE}{}^C H^{BE} = - W_{AE}{}^B H^{CE} \ .\label{MetricCompAppD}
\ee
\end{itemize}
Under these assumptions we will show here that the generalized Ricci
scalar equals the scalar potential of maximal supergravity. Let us
emphasize that we will not solve equations (\ref{TorsionlessAppD}) nor
(\ref{MetricCompAppD}), instead we will only use them as implicit
equations.

The Ricci scalar is defined as
\be
{\cal R} =  {\cal}H^{\bar A \bar B} {\cal R}_{\bar A \bar B}\ ,
\ee
and it is convenient to split it as
\be
{\cal R} = {\cal R}_0 - 4 H^{\bar A \bar B}\vartheta_{\bar A
}\vartheta_{\bar B} - H^{\bar A \bar B} \nabla_N V^N_{\bar A \bar B}\ ,
\label{R+R0+otros}
\ee
where
\be
{\cal R}_0 =  \frac 1 2 H^{\bar A \bar B} W_{\bar D \bar A}{}^{\bar E} ( 2
W_{\bar B \bar E}{}^{\bar D} + Y^{\bar D}{}_{\bar E}{}^{\bar F}{}_{\bar G}
W_{\bar F \bar B}{}^{\bar G}) - H^{\bar A \bar B} W_{\bar E\bar A}{}^{\bar
D} F_{\bar B \bar D}{}^{\bar E}\ .
\ee
In ${\cal R}_0$, the $Y$ can be decomposed as in (\ref{Y}) and then using
the torsionless condition (\ref{TorsionlessAppD}) together with the trace
of the spin connection (\ref{spintrace}) one obtains
\be
{\cal R}_0 = -\frac 1 2 H^{\bar A \bar B} W_{\bar E\bar A}{}^{\bar D}
F_{\bar B \bar D}{}^{\bar E} + \frac 1 2 H^{\bar A \bar B}(W_{\bar D \bar
A}{}^{\bar E}W_{\bar E \bar B}{}^{\bar D} + W_{\bar E \bar A }{}^{\bar D}
W_{\bar B \bar D}{}^{\bar E})\ ,
\ee
where last two terms vanish due to metric compatibility
(\ref{MetricCompAppD}). We can now use the decomposition of the fluxes $F$
as in (\ref{embedding tensor}), (\ref{gaugings912}) and (\ref{gaugings
56}) to obtain
\be
{\cal R}_0 = - \frac 1 2 H^{\bar A \bar B} W_{\bar D \bar A}{}^{\bar E}(
X_{\bar B \bar E}{}^{\bar D } + 8 P_{(adj)}{}^{\bar D}{}_{\bar E}{}^{\bar
F}{}_{\bar B} \vartheta_{\bar F} - \vartheta_{\bar B} \delta_{\bar
E}^{\bar D})\ .
\ee
When the projector acts on the spin connection, one can use again the
torsionless condition (\ref{TorsionlessAppD}) and metric compatibility
(\ref{MetricCompAppD}) to re-cast this expression in the form
\be
{\cal R}_0 = - \frac 1 2 H^{\bar A \bar B} \left(W_{\bar D \bar A}{}^{\bar
E} X_{\bar B \bar E}{}^{\bar D} + \frac 8 3 \vartheta_{\bar A}
\vartheta_{\bar B} - \frac 23 F_{\bar A \bar B}{}^{\bar F} \vartheta_{\bar
F}\right)\ .
\ee
Now plugging this in (\ref{R+R0+otros}) we get
\be
{\cal R } = - \frac 1 2 H^{\bar A \bar B} \left(W_{\bar D \bar A}{}^{\bar
E} X_{\bar B \bar E}{}^{\bar D} + \frac {32} 3 \vartheta_{\bar A}
\vartheta_{\bar B} - \frac 23 F_{\bar A \bar B}{}^{\bar F} \vartheta_{\bar
F}\right) -  H^{\bar A  \bar B} \nabla_N V^N{}_{(\bar A \bar B)}\ .
\ee
Here, the first term can be treated as follows. First, we use the fact
that the last two indices of $X$ project the corresponding indices of $W$
into the adjoint
\be
X_{\bar B \bar E}{}^{\bar D}\ W_{\bar D \bar A}{}^{\bar E} = X_{\bar B
\bar E}{}^{\bar D} \ P_{(adj)}{}^{\bar D}{}_{\bar E}{}^{\bar F}{}_{\bar
G}\ W_{\bar F \bar A}{}^{\bar G}\ ,
\ee
and using the torsionless condition (\ref{TorsionlessAppD}) one obtains
\be
-\frac 1 2 H^{\bar A \bar B} W_{\bar D \bar A}{}^{\bar E} X_{\bar B \bar
E}{}^{\bar D} = - \frac 14 H^{\bar A \bar B } (X_{\bar B \bar F }{}^{\bar
G} W_{\bar A \bar G}{}^{\bar F} - X_{\bar B \bar F }{}^{\bar G} F_{\bar A
\bar G}{}^{\bar F})\ .
 \ee
Finally, the first term here can be massaged by explicitly extracting a
projector to the $\bf 912$ from $X$
 \be
 H^{\bar A \bar B } \ X_{\bar B \bar F }{}^{\bar G}\  W_{\bar A \bar
G}{}^{\bar F} = H^{\bar A \bar B }\ P_{(912)\bar B \bar F}{}^{\bar
G},^{\bar M \bar N}{}_{\bar P} \ X_{\bar M \bar N}{}^{\bar P}\  W_{\bar A
\bar G}{}^{\bar F}\ ,
 \ee
  and exploiting the fact that the projector is invariant under rotations
with the generalized metric
\be
P_{(912)\bar A \bar B}{}^{\bar C},^{\bar M \bar N}{}_{\bar P} H^{\bar
A}{}_{\bar A'}H^{\bar B}{}_{\bar B'}H_{\bar C}{}^{\bar C'}H_{\bar
M}{}^{\bar M'}H_{\bar N}{}^{\bar N'}H^{\bar P}{}_{\bar P'} = P_{(912)\bar
A' \bar B'}{}^{\bar C'},^{\bar M' \bar N'}{}_{\bar P'}\ ,
\ee
after some algebra one obtains the final result
\bea
\frac 1 4 {\cal R} &=& \frac 1 {672} \left[{H}^{\bar A\bar D}{\bar
H}^{\bar B\bar E}{H}_{\bar C\bar F} X_{\bar A\bar B}{}^{\bar C}
X_{\bar D\bar E}{}^{\bar F} + 7 {H}^{\bar A\bar B} X_{\bar A\bar
C}{}^{\bar D} F_{\bar B\bar D}{}^{\bar C}\right]\nn\\
&& - \frac 4 3 H^{\bar A \bar B} \vartheta_{\bar A} \vartheta_{\bar B} +
\frac 1 {12} H^{\bar A\bar B} F_{\bar A \bar B}{}^{\bar F} \vartheta_{\bar
F} - \frac 1 4 H^{\bar A  \bar B} \nabla_N V^N{}_{(\bar A \bar B)}\ .
\eea

Being expressed purely in terms of fluxes, we see that the undetermined
pieces of the spin connection dropped out. Remarkably, this takes the
exact same form as the scalar potential of gauged
maximal supergravity \cite{de Wit:2007mt} if we identify the generalized
metric
with the moduli space metric ${\cal M}$ and take the gaugings in the $\bf
56$ to vanish
\be
\frac 1 4 {\cal R} = \frac 1 {672} \left[{H}^{\bar A\bar D}{\bar H}^{\bar
B\bar E}{H}_{\bar C\bar F}\ X_{\bar A\bar B}{}^{\bar C}\
X_{\bar D\bar E}{}^{\bar F} + 7 {H}^{\bar A\bar B}\ X_{\bar A\bar
C}{}^{\bar D}\ X_{\bar B\bar D}{}^{\bar C}\right]\ .
\ee
Note also that this is true for any
torsionless and metric-compatible connection, and the concrete expression
of the
determined part does not need to be known. In fact, we never needed to
solve for
the spin connection, but only used the equation that defines it
implicitly.

Finally, let us mention that the factor $7$ in the scalar potential comes
form a projection of the spin connection into the space of fluxes
(\ref{projsSpin}). It is known that it is fixed by supersymmetry \cite{de
Wit:2007mt}, so one can wonder where does supersymmetry arise in all this
analysis. This factor is actually $1/a_1$ in the language of Appendix
\ref{AppC}, and it is set by supersymmetry in an indirect way. As we
explained in Appendix \ref{AppC}, the projectors to the irreducible reps
of the direct product of the fundamental and the adjoint are normalized to
add up to unity. Since supersymmetry projects out some reps through a
linear constraint, the remaining are normalized in such a way that they
capture information about supersymmetry and this is exactly how this
coefficient is obtained here.

\end{document}